\documentclass[pdflatex,sn-mathphys-num]{sn-jnl}%
\usepackage[T1]{fontenc}

\usepackage{graphicx}%
\usepackage{multirow}%
\usepackage{amsmath,amssymb,amsfonts}%
\usepackage{amsthm}%
\usepackage{mathrsfs}%
\usepackage[title]{appendix}%
\usepackage{xcolor}%
\usepackage{textcomp}%
\usepackage{manyfoot}%
\usepackage{booktabs}%
\usepackage{algorithm}%
\usepackage{algorithmicx}%
\usepackage{algpseudocode}%
\usepackage{listings}%
\usepackage{threeparttable}
\usepackage{orcidlink}

\theoremstyle{thmstyleone}%

\theoremstyle{thmstyletwo}%

\theoremstyle{thmstylethree}%

\newcommand{\bigO}{\mathcal{O}}

\begin{document}

\title{Quantum Internet: Technologies, Protocols, and Research Challenges}

\author*[1,2]{\fnm{Vinay} \sur{Kumar}\orcidlink{0000-0002-4635-3237}}\email{vinay.kumar@phd.unipi.it}

\author[2]{\fnm{Claudio} \sur{Cicconetti}\orcidlink{0000-0003-4503-4223}}\email{c.cicconetti@iit.cnr.it}

\author[2]{\fnm{Marco} \sur{Conti}\orcidlink{0000-0003-4097-4064}}\email{marco.conti@iit.cnr.it}

\author[2]{\fnm{Andrea} \sur{Passarella}\orcidlink{0000-0002-1694-612X}}\email{a.passarella@iit.cnr.it}

\affil[1]{\orgdiv{Department of Information Engineering}, \orgname{University of Pisa}, \orgaddress{\street{Via G. Caruso 16}, \city{Pisa}, \postcode{56122}, \state{Pisa}, \country{Italy}}}

\affil[2]{\orgdiv{Institute for Informatics and Telematics (IIT)}, \orgname{National Research Council (CNR)}, \orgaddress{\street{Via Moruzzi 1}, \city{Pisa}, \postcode{56126}, \state{Pisa}, \country{Italy}}}

\abstract{
As the field of the quantum internet advances, a comprehensive guide to navigate its complexities has become increasingly crucial. While quantum computing shares foundational principles with the quantum internet, distinguishing between the two is essential for further development and deeper understanding. This work systematically introduces the quantum internet by discussing its importance, core components, operational mechanisms, anticipated timeline for viability, key contributors, major challenges, and future directions. Additionally, it presents the fundamental concepts of quantum mechanics that underpin the technology, offering a clear and targeted overview intended for researchers and industry professionals and laying the groundwork for future innovations and research in the field.

}


\maketitle

\vspace{-0.5em}
\noindent\textbf{Note.} This is the author accepted manuscript. \\
\noindent\textbf{Please cite as}: Kumar, V., Cicconetti, C., Conti, M. et al. 
Quantum Internet: Technologies, Protocols, and Research Challenges. 
\textit{Int J Netw Distrib Comput} 13, 22 (2025). 
\href{https://doi.org/10.1007/s44227-025-00060-5}{https://doi.org/10.1007/s44227-025-00060-5}

\vfill 

\section{Introduction}\label{sec:intro}
The quantum internet\footnote{
In current literature, ``quantum internet" and ``quantum networks" are often used interchangeably due to the nascent state of the field. In this paper, however, when we use ``quantum internet" we refer to its broader vision, impact, and potential applications, whereas ``quantum networks" denotes the specific protocols and operational mechanisms. It is important to note that today, only the concept of a quantum internet exists, and almost all studies refer to quantum networks.} has garnered increasing attention as quantum computing technology starts emerging in the market. This groundbreaking technology holds the potential to revolutionise security \cite{nurhadi18, mehic20, cao22, bernstein17, broadbent09, fitzsimons17, ben05, vaccaro07, hahn20, murta20, broadbent20, ganz17, rahman19, rozenman23}, computing \cite{cacciapuoti19, caleffi24, cuomo20}, and specialised applications such as sensing \cite{degen17} and time synchronisation \cite{chuang20} by leveraging fundamental quantum phenomena. The concept of integrating the quantum internet alongside the classical internet has resonated with the research community, driving efforts to explore ways to capitalise on the principles of quantum mechanics. This interest has sparked research into the architecture of quantum networks, the identification of potential applications and use cases, efficient entanglement distribution\footnote{There are two methods for distributing entanglement. The first involves directly sending one of the qubits of an entangled pair to the target location. The second method utilises entanglement swapping to distribute entangled pairs to two endpoints. Directly sending entangled pairs to the target is generally not advisable. Therefore, in this work, when we refer to ``distribution," we specifically mean the use of the entanglement swapping procedure to distribute the entanglement.} within quantum networks and the development of test beds utilising various physical systems \cite{pompili21, van22, luo22, krutyanskiy23}.

While physical quantum computing devices are essential for the global quantum internet, some early applications do not require fault-tolerant quantum devices and are simpler to implement (refer Section~\ref{ssec:stages}). Significant progress has been made in the early stages of quantum internet, particularly in the development of quantum key distribution (QKD)-based networks, both terrestrial \cite{cao22} and satellite-based \cite{liao17, liao18, chen21km4600}. These developments are elaborated upon in Section~\ref{ssec:test_beds} and Section~\ref{sec:current_initiatives}. Steady advancements are evident in the enhancement and deployment of QKD-based networks. At the same time, later stages of the quantum internet, particularly those that rely on quantum entanglement, have also seen progress in various areas. Notably, one of the critical aspects—end-to-end entanglement distribution, a critical aspect that will be discussed in more detail in Section~\ref{ssec:basics}—has become a focus of considerable research in recent years. 

The core concept of the quantum internet—harnessing quantum mechanical phenomena to enable certain applications—comes with inherent limitations dictated by the laws of quantum physics, such as decoherence \cite{schlosshauer19} and the no-cloning theorem \cite{wootters82}, both of which are introduced in Section~\ref{ssec:basics}. These properties distinguish the quantum internet from the classical internet. The challenge lies in the innovative use of quantum phenomena to develop quantum applications that offer advantages over classical methods while adhering to these quantum mechanical restrictions. Advancing the vision of a quantum internet requires overcoming several technical hurdles, many of which overlap with the field of quantum computing. These include the development of quantum memories \cite{bussieres13, heshami16} and the realisation of quantum error correction \cite{roffe19} to ensure robustness against errors. 

\begin{figure}
\centering
\includegraphics[width=\columnwidth]{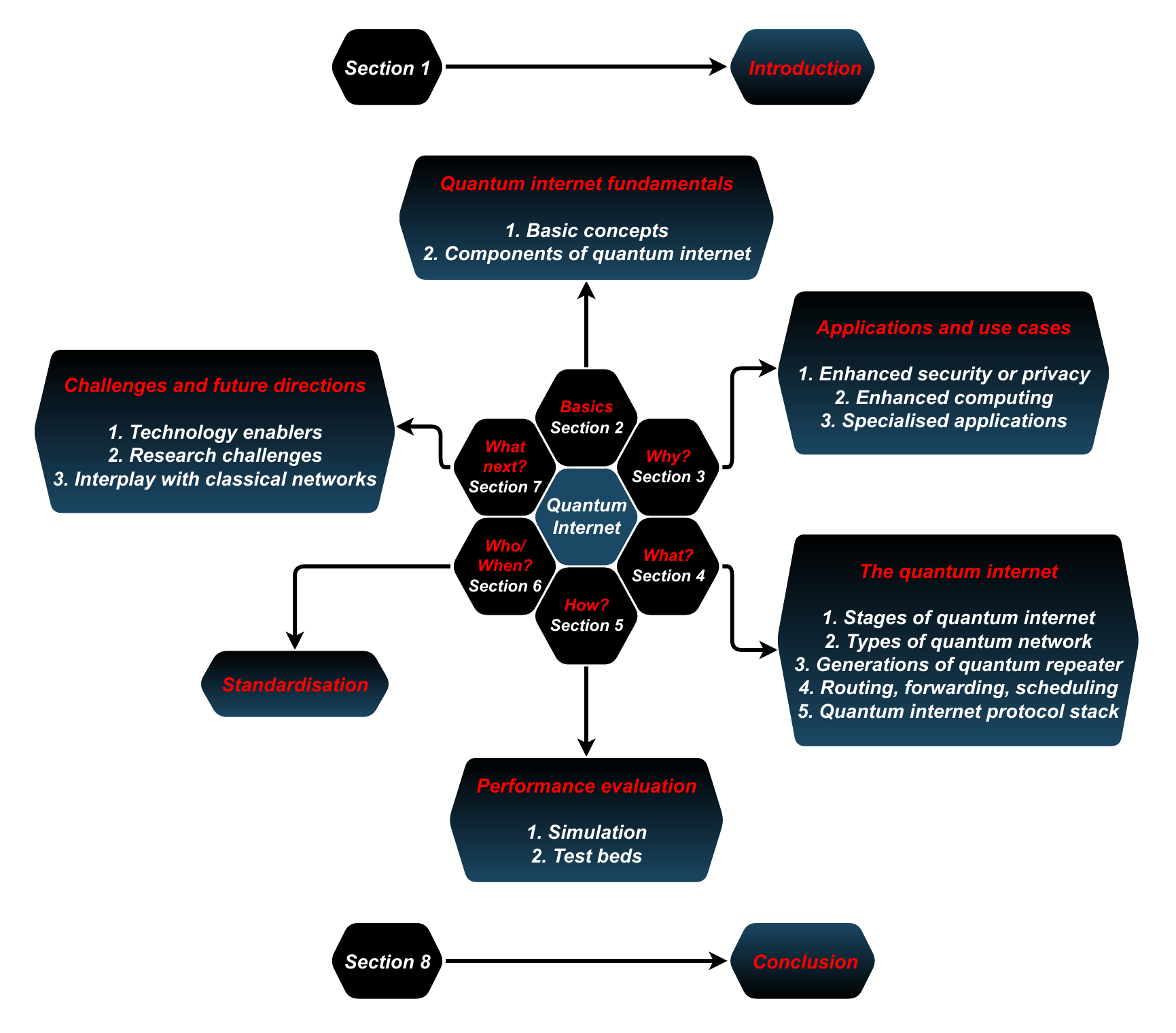}
\caption{Table of contents}
    \label{fig:intro_blocks}
\end{figure}

As advancements in quantum technology continue to emerge, adapting the development of the quantum internet to these evolving innovations will be a dynamic challenge. However, establishing a quantum internet based on currently understood technological parameters is a problem that can be addressed today. With this in mind, we aim to provide an introduction to the quantum internet to inspire further research alongside quantum computing.
Given that the quantum internet shares fundamental principles with quantum computing, this work systematically introduces the field while maintaining a focus specifically on the quantum internet. As this is a novel area, research has progressed on various fronts. To support further exploration, we also provide a comprehensive list of relevant review articles on each topic discussed, as summarised in table~\ref{tab:list_review}. While each of them provides a thorough view on a specific topic which is relevant for the quantum internet, none of them takes a holistic perspective on the topic, which is the main goal of this survey.

\begin{table}[h!]
\centering
\caption{A list of relevant review articles to each topic of this article.}
\label{tab:list_review}
\begin{tabular}{p{5cm}p{2.6cm}p{3.6cm}}
\toprule
Topic & This paper & Additional Readings \\
\midrule
Entanglement Purification & Section~\ref{ssec:basics} & \cite{yan23} \\
Stages of Quantum Internet & Section~\ref{ssec:stages} &  \cite{wehner18} \\
Entanglement Routing & Section~\ref{ssec:routing_forwarding_scheduling} & \cite{dupuy23, abane24} \\
Quantum Internet protocol stack & Section~\ref{ssec:protocol_stack} &  \cite{illiano22, li24} \\
Simulation Tools & Section~\ref{ssec:simulation} &  \cite{bel24} \\
QKD networks & Section~\ref{ssec:test_beds} &  \cite{mehic20, cao22} \\
\bottomrule
\end{tabular}
\end{table}

The organisation of the paper is as follows and is also illustrated in Figure~\ref{fig:intro_blocks}. In Section~\ref{sec:qi_fundamentals}, we establish the fundamentals of the quantum internet, including basic concepts and key components of the quantum internet. In Section~\ref{sec:application}, we discuss the applications and use cases. In Section~\ref{sec:architecture}, we provide an in-depth architecture of the quantum internet. In Section~\ref{sec:performance_evaluation}, we discuss the current performance evaluation methods, including simulation tools, models, and test beds. In Section~\ref{sec:current_initiatives}, we discuss the current initiatives and collaborations, particularly concerning standardisation efforts. In Section~\ref{sec:challenges_and_future}, we give the technical hurdles and future research directions. At last, Section~\ref{sec:conclusion} concludes the paper.

\section{Quantum Internet Fundamentals}\label{sec:qi_fundamentals}
The quantum internet is fundamentally different from the classical internet because of the acquired properties from the quantum mechanical phenomena. These acquired counterintuitive properties in the classical world provide the backbone of the quantum internet. In this Section, we explain the basic concepts relevant to the quantum internet, followed by a component-by-component explanation of the quantum internet.

\subsection{Basic Concepts}\label{ssec:basics}

We now introduce the basic terminology and definitions for quantum computing, which will be useful throughout the paper to readers unfamiliar with the topic.

\textbf{Qubit:} A qubit\footnote{While a qubit is a two-level quantum system, a generalised version with
n-levels is known as a \textit{qudit}. Despite having interesting properties that could be exploited for enhanced systems or applications, qudit technologies lag significantly behind their equivalent, that is, qubits \cite{wang20}. Therefore, in this work, we will not consider qudits.} is fundamental to quantum computing, similar to a bit in classical computing. While a classical bit on a normal computer can be in a state of 0 or 1, the state of a quantum bit is defined by:
    \begin{equation}\label{eq:sample_psi}
        |\psi \rangle = \alpha |0 \rangle + \beta |1 \rangle
    \end{equation}
    The state $|\psi \rangle $ is said to be in a linear combination of states or a \textit{superposition state} which collapses to 0 or 1 upon \textit{measurement} with a probability of $|\alpha|^2$ and $|\beta|^2$ respectively depending upon the \textit{probability amplitudes} that is, $\alpha$ and $\beta$, which are complex coefficients. Hence by definition $| \alpha |^2 + |\beta|^2 = 1$. 
    
The notation $|\psi \rangle$ used here is a standard for representing quantum mechanical states called \textit{bra-ket} notation introduced by Paul Dirac. Here the represented $|\psi \rangle$ is a \textit{ket} 
which is a vector in a complex vector space called \textit{Hilbert space} $\mathcal{H}$. A Hilbert space is a linear vector space with three additional properties that is, strictly positive scalar product, separability, and completeness \cite{zettili09}. For every ket $| \psi \rangle $, there exists a unique \textit{bra} $\langle \psi |$, which belongs to the corresponding \textit{dual-Hilbert space} $\mathcal{H}_d$. While the eq~\ref{eq:sample_psi} above describes a single qubit system, multi-qubit systems can be represented by extending this notation. In general, an \( n \)-qubit system is described by:
\begin{equation}
|\psi \rangle = \sum_{i=0}^{2^n-1} c_i\, |i\rangle,
\end{equation}
where \( |i\rangle \) represents the computational basis states corresponding to the binary representation of \( i \), and \( c_i \in \mathbb{C} \) are the complex probability amplitudes satisfying the normalization condition:
\begin{equation}
\sum_{i=0}^{2^n-1} |c_i|^2 = 1.
\end{equation}
For example, a two-qubit system is expressed as:
\begin{equation}
    |\psi \rangle = \alpha |00 \rangle + \beta |01 \rangle + \gamma |10 \rangle + \delta |11 \rangle
\end{equation} 
where $|\alpha|^2+|\beta|^2+|\gamma|^2+|\delta|^2=1$.

\vspace*{0.5cm}

\textbf{Quantum gates:} A quantum gate is a fundamental building block of quantum circuits that can manipulate a quantum state or qubit. Quantum gates, for example, $X$, Hadamard, Toffoli, $CZ$, etc., can be considered analogous to logic gates like AND, OR, XOR, etc, in classical computing. In quantum mechanical terms, a quantum gate is an operator applied to a ket $|\psi \rangle = \alpha |0 \rangle + \beta |1 \rangle$ that changes it to a ket $|\psi^{'} \rangle = \alpha^{'} |0 \rangle + \beta^{'} |1 \rangle$. The quantum gates can be represented by square matrices which act linearly on the quantum states \cite{nielsen10}. For example, the quantum NOT gate that is, $X$ is expressed as follows:
    \begin{equation}\label{eq:x_gate}
     X = \begin{pmatrix}
     0 & 1 \\
     1 & 0 
     \end{pmatrix},
    \end{equation}
and it has the same effect as its classical counterpart: the classical NOT gate flips a $0$ into a $1$ (and a $1$ into a $0$), whereas the quantum X gate flips a $|0\rangle$ into a $|1\rangle$ (and a $|1\rangle$ into a $|0\rangle$).
For a matrix $U$ to be a quantum gate, it must be unitary; that is, it should satisfy the condition $U^{\dagger}U = I$. Here, the notation $U^{\dagger}$ is called the \textit{adjoint}, that is, the transpose conjugate of the matrix $U$. This is necessary to maintain the sum of probabilities to be one after the operation of the quantum gate, that is, $| \alpha^{'} |^2 + |\beta^{'}|^2 = 1$.
    
A quantum gate can be categorised as a $n$-qubit gate where $n$ is the number of qubits the gate can be operated on. The $X$ gate represented in eq~\ref{eq:x_gate} is a 1-qubit gate. An example of a 2-qubit gate is $CZ$, which is represented as follows:
    \begin{equation}\label{eq:cz_gate}
        CZ = \begin{pmatrix}
         1 & 0 & 0 & 0 \\
         0 & 1 & 0 & 0 \\
         0 & 0 & 1 & 0 \\
         0 & 0 & 0 & -1
        \end{pmatrix}.
    \end{equation}
    In general, a $n$-qubit quantum gate corresponds to a matrix of dimension of $2^n$.

\vspace*{0.5cm}

\textbf{Entanglement:} Entanglement is an important quantum phenomenon that is core to the quantum internet. Till now, we have been introducing concepts with an analogy to its classical counterparts. But from now on, that approach would not be possible as the following concepts do not have a classical counterpart. Entanglement refers to a system prepared so that there are correlations among quantum bits or states independent of distance, famously dubbed by Einstein as \textit{spooky action at a distance}. In principle, entanglement can be established between any number of qubits known as Greenberger–Horne–Zeilinger (\textit{GHZ-state}) \cite{greenberger89, greenberger90} and \textit{W-state} \cite{dur00} or even in multiple degrees of freedoms known as hyperentanglement\footnote{Due to the complexity of managing multiple degrees of freedom, the use of hyperentanglement for entanglement distribution within the quantum internet is rarely addressed. However, protocols involving teleportation, swapping, and purification—which could leverage hyperentanglement—are well-documented in the literature and remain an active area of research. For further reading on this topic, a relevant review article is available at \cite{deng17}.} \cite{kwiat97} as shown in figure~\ref{fig:entanglement_type}.
But for the sake of the introduction of the concept, let's consider the case of a two-qubit system with a single degree of freedom.
Throughout this paper, our discussion will concentrate on the workings of the quantum internet using bipartite entanglement, which has been a primary focus in the literature. However, multi-partite-based entanglement distribution is discussed briefly in Section~\ref{ssec:future_works}.
In particular, the quantum internet is concerned with the distribution of a special two-qubit entangled system, named \textit{EPR pair} \cite{einstein35} or \textit{Bell state} \cite{bell64}. An example of Bell state is as follows:

\begin{figure}
\centering
\includegraphics[width=\columnwidth]{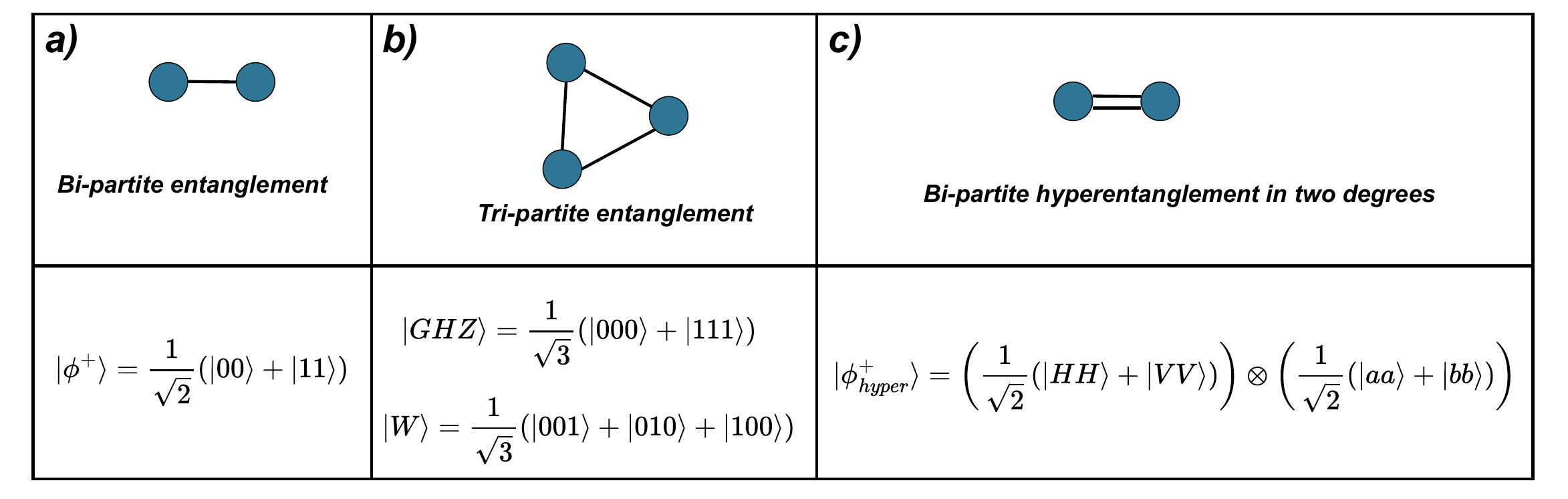}
\caption{Examples of different types of entanglement. \textbf{a)} a bipartite entangled state depicted by the Bell state; \textbf{b)} two forms of tripartite entanglement, one following the GHZ state and the other the W state; \textbf{c)} a bipartite hyperentanglement in two degrees of freedom, combining polarization and spatial modes.}
    \label{fig:entanglement_type}
\end{figure}

    \begin{equation}\label{eq:bell_state}
        | \phi^+ \rangle = \frac{1}{\sqrt{2}} \left( |00 \rangle + |11 \rangle \right).
    \end{equation}
Upon measurement of the two-qubit state $|\phi^+ \rangle$, the whole system collapses to $|00 \rangle$ or $|11 \rangle$ with a 50\% probability. Independent of the state the system collapses to, the two-qubit measurement result is correlated: it is impossible to know beforehand the result of the measurement, but if the measurement corresponding to the first qubit is $0$, then the second qubit measurement will be $0$, and the same happens if the first qubit is measured as a $1$, in which case the second qubit will also be measured as $1$.
    
The other Bell states are as follows:
        \begin{equation}\label{eq:other_bell_states}
        \begin{split}
            | \phi^- \rangle = \frac{1}{\sqrt{2}} \left( |00 \rangle - |11 \rangle \right) \\
            | \psi^+ \rangle = \frac{1}{\sqrt{2}} \left( |01 \rangle + |10 \rangle \right) \\
            | \psi^- \rangle = \frac{1}{\sqrt{2}} \left( |01 \rangle - |10 \rangle \right)
        \end{split}.
    \end{equation}

The \( + \) and \( - \) symbols in the Bell states indicate the relative phase between the two basis states that make up each Bell state. Bell states are important mainly for two reasons.
First, they are maximally entangled, which means informally that it is not possible to prepare any quantum state where the entanglement between two qubits is stronger.
The formal definition of this property would require introducing additional and unnecessary notation.
Therefore, we suggest interested readers to seek further information in textbooks, such as~\cite{zettili09, nielsen10}.
Second, it is possible to transform a Bell state into any arbitrary state of choice via local operations only, that is, the application of quantum gates.
In summary, while one could distribute arbitrary quantum states in the quantum internet, doing so with Bell pairs only is an efficient and practical alternative, which is universally accepted by the research community.

\vspace*{0.5cm}

\textbf{Decoherence:} In quantum information, \emph{decoherence} refers to the gradual loss of ``quantum-ness" of a quantum state, usually caused by its interaction with the environment. This disruption occurs because the quantum state becomes entangled with its surroundings, which ``measures" or interacts with it in a way that scrambles its coherence. Over time or due to specific interactions, such as qubit measurements or quantum gates, this loss of coherence leads the quantum state to behave more like a classical system, losing the features that made it quantum \cite{schlosshauer19}.

For the quantum internet, we can think of decoherence as the natural degradation of quantum information as it travels or undergoes operations. This loss of coherence means that information can become noisy or unusable, and it places limits on how long and how far we can reliably use a quantum state for communication or computation. Addressing and minimizing decoherence is essential to maintaining the quality of information in quantum internet, as even small disturbances from the environment can lead to a breakdown in the system’s ability to preserve entanglement and other quantum correlations.

\vspace*{0.5cm}

\textbf{Fidelity:} It is a metric which quantifies the closeness of two quantum states. Let $|\psi_i \rangle$ and $|\psi_j \rangle$ be two quantum states then the fidelity $F_{ij}$ of these states is given by:
    \begin{equation}\label{eq:fidelity_bra_ket}
        F_{ij} = | \langle \psi_i | \psi_j \rangle|^2
    \end{equation}
where $\langle \psi_i | \psi_j \rangle$ is the inner product between the states. For pure quantum states—states that are fully coherent and not mixed with any probability distribution over different possibilities—it is $0 \leq F_{ij} \leq 1$, where $F_{ij} = 0$ means that the states are orthogonal to each other and $F_{ij} = 1$ means that the states are equivalent.

\vspace*{0.5cm}

\textbf{No cloning theorem:} The no-cloning theorem states that it is not possible to copy an unknown quantum state. More formally, there does not exist a universal quantum operation (unitary transformation) that takes any arbitrary quantum state $|\psi \rangle $ along with a standard ``blank'' state $|e \rangle$ and produces two copies of $|\psi \rangle$, that is, there is no unitary operation $U$ such that for all $|\psi \rangle$ \cite{wootters82},
    \begin{equation}
        U(|\psi\rangle \otimes |e\rangle) = |\psi\rangle \otimes |\psi\rangle
    \end{equation}

This theorem is one of the fundamental concepts that distinguishes the quantum internet from the classical internet. In classical systems, copying information is extensively utilized for various purposes, including assisting data transmission, error correction, caching, load balancing, and ensuring backup \& redundancy.

\vspace*{0.5cm}

\textbf{Teleportation:} Teleportation of a quantum state refers to sending an arbitrary quantum state to an arbitrary distance using quantum correlations \cite{bennett93}. In Figure~\ref{fig:teleport_swap_chain}a Alice wants to send an arbitrary quantum state $|\psi \rangle_0^{A_1} = \alpha |0 \rangle + \beta |1 \rangle$ to Bob. Let Alice and Bob initially have an EPR pair $| \phi^+ \rangle^{A_2 B}$ shared among them. Then, the state of the collective three-qubit system can be written as follows:
\begin{figure}
\centering
\includegraphics[width=\columnwidth]{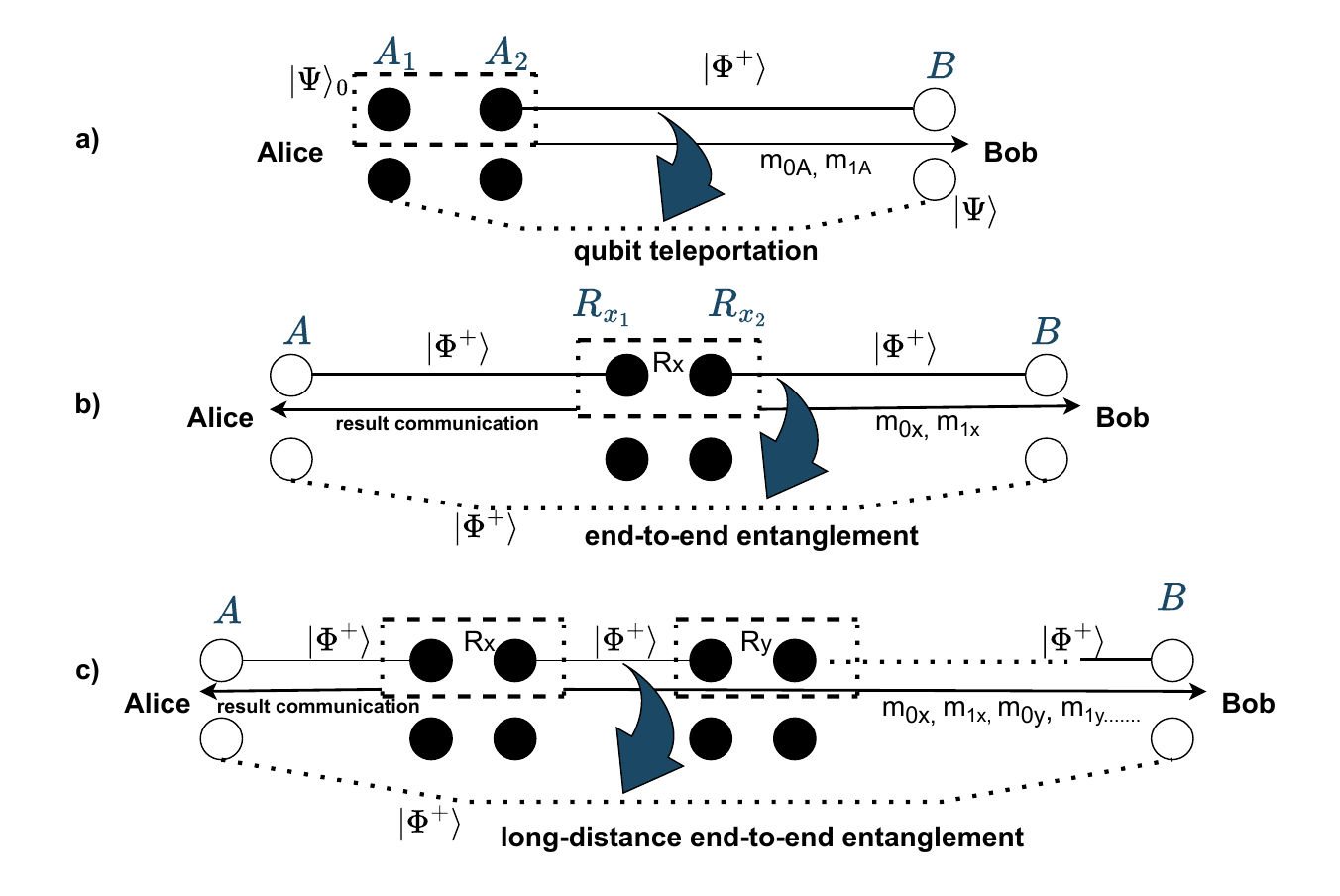}
\caption{\textbf{a) Teleportation:} Teleportation of Alice's qubit \(A_1\) from state \(|\Psi\rangle_0\) to Bob's qubit \(B\) (yielding state \(|\Psi\rangle\) with fidelity \( |\langle \Psi_0 | \Psi \rangle|^2 \)) is achieved via a Bell state measurement on qubits \(A_1\) and \(A_2\). Alice then communicates the resulting classical bits (\(m_{0A}\) and \(m_{1A}\)) to Bob, who applies the appropriate quantum operations on \(B\) based on these outcomes. \textbf{b) Entanglement Swapping:} Entanglement swapping between Alice and Bob is achieved by converting two bipartite entangled pairs—one between Alice and repeater \(R_x\) (qubits \(A\) and \(R_{x_1}\)) and one between repeater \(R_x\) and Bob (qubits \(R_{x_2}\) and \(B\))—into a direct (end-to-end) entanglement between Alice and Bob (qubits \(A\) and \(B\)). This process is executed by performing a Bell state measurement on \(R_x\)'s qubits \(R_{x_1}\) and \(R_{x_2}\), communicating the resulting classical bits (\(m_{0A}\) and \(m_{1A}\)) to Bob, who then applies the appropriate quantum operations on his qubit \(B\). \textbf{c) A linear network:} End-to-end bipartite entanglement between two distant end nodes, Alice and Bob (qubits \(A\) and \(B\)), is established via a cascade of entanglement swapping along a linear chain of quantum repeaters. At each repeater, a Bell state measurement is performed and the corresponding classical bits (\(m_{0i}\) and \(m_{1i}\)) from the \(i^{th}\) quantum repeater are communicated to Bob, who applies the appropriate quantum operations on his qubit \(B\). (modified from \cite{vk24}).}
    \label{fig:teleport_swap_chain}
\end{figure}
    \begin{equation}
        | \psi \rangle_0^{A_1} | \phi^+ \rangle^{A_2 B} = \left( \alpha |0 \rangle + \beta |1 \rangle \right)^{A_1} \ \frac{1}{\sqrt{2}} \left( |00 \rangle + |11 \rangle \right)^{A_2 B}
    \end{equation}
    \begin{equation}
        \Rightarrow | \psi \rangle_0^{A_1} | \phi^+ \rangle^{A_2 B} = \frac{1}{\sqrt{2}} (\alpha |000 \rangle + \alpha |011 \rangle + \beta | 100 \rangle + \beta|111\rangle)^{A_1 A_2 B}
    \end{equation}
Using the definitions of the Bell states in eq.~\ref{eq:bell_state} and \ref{eq:other_bell_states} we have:
\begin{equation}
\begin{split}
    \Rightarrow | \psi \rangle_0^{A_1} | \phi^+ \rangle^{A_2 B} = \frac{1}{2} (\alpha (|\phi^+ \rangle + |\phi^- \rangle) |0 \rangle + \alpha (|\psi^+ \rangle + |\psi^- \rangle) | 1 \rangle + \\
    \beta (|\psi^+\rangle -|\psi^- \rangle)|0\rangle + \beta (|\phi^+ \rangle - |\phi^-\rangle)|1\rangle)^{A_1 A_2 B}
    \end{split}
\end{equation}
    \begin{equation} \label{eq:teleport_expanded}
    \begin{split}
        \Rightarrow | \psi \rangle_0^{A_1} | \phi^+ \rangle^{A_2 B} = \frac{1}{2} | \phi^+ \rangle^{A_1 A_2} \left( \alpha |0 \rangle + \beta |1  \rangle \right)^B + \frac{1}{2} | \phi^- \rangle^{A_1 A_2} \left( \alpha |0 \rangle - \beta |1  \rangle \right)^B \\
        + \frac{1}{2} | \psi^+ \rangle^{A_1 A_2} \left( \alpha |1 \rangle + \beta |0  \rangle \right)^B + \frac{1}{2} | \psi^- \rangle^{A_1 A_2} \left( \alpha |1 \rangle - \beta |0  \rangle \right)^B
    \end{split}
    \end{equation}
where,
    \begin{equation}\label{eq:bell_basis}
        \begin{split}
            | \phi^+ \rangle = \frac{1}{\sqrt{2}} \left( |00 \rangle + |11 \rangle \right) \\
            | \phi^- \rangle = \frac{1}{\sqrt{2}} \left( |00 \rangle - |11 \rangle \right) \\
            | \psi^+ \rangle = \frac{1}{\sqrt{2}} \left( |01 \rangle + |10 \rangle \right) \\
            | \psi^- \rangle = \frac{1}{\sqrt{2}} \left( |01 \rangle - |10 \rangle \right) 
        \end{split}
    \end{equation}
From eq~\ref{eq:teleport_expanded}, a simultaneous bell state measurement by Alice that is, on qubits $A_1$ and $A_2$ would result in either of four bell states that is, $|\phi^+ \rangle^{A_1 A_2}, |\phi^- \rangle^{A_1 A_2}, |\psi^+ \rangle^{A_1 A_2}$, and $|\psi^- \rangle^{A_1 A_2}$. Meanwhile the corresponding state of Bob's qubit that is, $B$ would be in the state: $\left( \alpha |0 \rangle + \beta |1 \rangle \right)^B, \left( \alpha |0 \rangle - \beta |1 \rangle \right)^B, \left( \alpha |1 \rangle + \beta |0 \rangle \right)^B$, and $\left( \alpha |1 \rangle - \beta |0 \rangle \right)^B$. The measurement results obtained by Alice are communicated to Bob via a classical channel. Depending upon these measurement results Bob would perform a single-qubit gate that is, $I$ for $| \phi^+ \rangle$, $\sigma_z$ for $|\phi^- \rangle$, $\sigma_x$ for $| \psi^+ \rangle$, and $\sigma_x \sigma_z$ for $|\psi^- \rangle$ which gets the state of Bob's qubit transformed to the state that was intended to teleport by Alice \cite{Weinfurter94}. This concludes the teleportation procedure as summarized in table~\ref{tab:bell_state_analysis}.
The net effect is that the source quantum state has been transferred to a remote party through the consumption of a Bell state shared by the parties and with the collapse of the origin state.
Note that this procedure does not violate speed-of-light constraints on the transfer of matter or information, as it relies on both the pre-distribution of a Bell state between Alice and Bob and the transmission of measurement results via a classical communication channel.

\begin{table}[h!]
\centering
\caption{\textbf{Bell state analysis for Teleportation:} Bell state measurement at Alice's qubits $A_1$ and $A_2$ \textit{\textbf{(column: $A_1A_2$ results)}} collapses (projects) the Bob's qubit $B$ to several different states \textit{\textbf{(column: $B$ result)}}. Hence, a single-qubit gate(s) is applied at Bob's qubit $B$ \textit{\textbf{(column: applied single-qubit gate at Bob ($B$))}} to arrive at correct final state of Bob's qubit $B$ \textit{\textbf{(column: final state at Bob ($B$))}}.}
\label{tab:bell_state_analysis}
\begin{tabular}{|l|l|l|l|}
\hline
$A_1 A_2$ results & $B$ result & applied single-qubit gate at Bob ($B$) & final state at Bob ($B$) \\
\hline
$|\phi^+ \rangle^{A_1 A_2}$ & $\left( \alpha |0 \rangle + \beta |1 \rangle \right)^B$ & $I$ & $\left( \alpha |0 \rangle + \beta |1 \rangle \right)^B$ \\
$|\phi^- \rangle^{A_1 A_2}$ & $\left( \alpha |0 \rangle - \beta |1 \rangle \right)^B$ & $\sigma_z$ & $\left( \alpha |0 \rangle + \beta |1 \rangle \right)^B$ \\
$|\psi^+ \rangle^{A_1 A_2}$ & $\left( \alpha |1 \rangle + \beta |0 \rangle \right)^B$ & $\sigma_x$ & $\left( \alpha |0 \rangle + \beta |1 \rangle \right)^B$ \\
$|\psi^- \rangle^{A_1 A_2}$ & $\left( \alpha |1 \rangle - \beta |0 \rangle \right)^B$ & $\sigma_x \sigma_z$ & $\left( \alpha |0 \rangle + \beta |1 \rangle \right)^B$ \\
\hline
\end{tabular}
\end{table}
\vspace*{0.5cm}
\textbf{Entanglement Swapping:} Entanglement swapping can be considered an extended version of the teleportation procedure above: entanglement swapping essentially \textit{swaps} two EPR pairs distributed at a shorter distance with a single EPR distributed at a longer distance \cite{zukowski93}. In contrast, the teleportation procedure is used to teleport an arbitrary quantum state, as shown above. In Figure~\ref{fig:teleport_swap_chain}b, two EPR pairs $|\phi^+ \rangle$ are initially distributed over three stations, that is, Alice, Repeater $R_x$, and Bob. Then, the state of the collective four-qubit system can be written as follows:
    \begin{equation}
        |\phi^+ \rangle^{A R_{x_1}} |\phi^+ \rangle^{R_{x_2} B} = \frac{1}{\sqrt{2}} \left( |00 \rangle + |11 \rangle \right)^{A R_{x_1}} \ \frac{1}{\sqrt{2}} \left( |00 \rangle + |11 \rangle \right)^{R_{x_2} B}
    \end{equation}
    \begin{equation}
        \Rightarrow |\phi^+ \rangle^{A R_{x_1}} |\phi^+ \rangle^{R_{x_2} B} = \frac{1}{2} (|0000 \rangle + |0011 \rangle + |1100 \rangle + |1111 \rangle)^{A R_{x_1} R_{x_2} B}
    \end{equation}
    Similar to teleportation, using the definitions of the Bell states in eq.~\ref{eq:bell_state} and \ref{eq:other_bell_states}, we have:
    \begin{equation}
        \begin{split}
            \Rightarrow |\phi^+ \rangle^{A R_{x_1}} |\phi^+ \rangle^{R_{x_2} B} = \frac{1}{2 \sqrt{2}} \left[ |0\rangle (|\phi^+ \rangle + |\phi^- \rangle) |0 \rangle + |0 \rangle (| \psi^+ \rangle +| \psi^- \rangle)|1 \rangle + \right. \\
            \left. |1 \rangle (|\psi^+ \rangle - |\psi^- \rangle) |0 \rangle + |1 \rangle (|\phi^+ \rangle - |\phi^- \rangle)|1 \rangle \right]^{A R_{x_1} R_{x_2} B}
        \end{split}
    \end{equation}

        \begin{equation}
        \begin{split}
            \Rightarrow |\phi^+ \rangle^{A R_{x_1}} |\phi^+ \rangle^{R_{x_2} B} = \frac{1}{2 \sqrt{2}} \left[ (|\phi^+ \rangle + |\phi^- \rangle) |00 \rangle + (| \psi^+ \rangle +| \psi^- \rangle)|01 \rangle + \right. \\
            \left. (|\psi^+ \rangle - |\psi^- \rangle) |10 \rangle + (|\phi^+ \rangle - |\phi^- \rangle)|11 \rangle \right]^{ R_{x_1} R_{x_2} A B}
        \end{split}
    \end{equation}
\begin{equation}\label{eq:swap_expanded}
\begin{split}
    \Rightarrow |\phi^+ \rangle^{A R_{x_1}} |\phi^+ \rangle^{R_{x_2} B} = \frac{1}{2} \left[
    | \phi^+ \rangle^{R_{x_1} R_{x_2}} \frac{1}{\sqrt{2}}\left( |00 \rangle + |11 \rangle \right)^{A B} \right. \\
    \left. + | \phi^- \rangle^{R_{x_1} R_{x_2}} \frac{1}{\sqrt{2}}\left( |00 \rangle - |11 \rangle \right)^{A B} \right. \\
    \left. + | \psi^+ \rangle^{R_{x_1} R_{x_2}} \frac{1}{\sqrt{2}}\left( |01 \rangle + |10 \rangle \right)^{A B} \right. \\
    \left. + | \psi^- \rangle^{R_{x_1} R_{x_2}} \frac{1}{\sqrt{2}}\left( |01 \rangle - |10 \rangle \right)^{A B} \right]
\end{split}
\end{equation}
where the Bell basis are given by eq~\ref{eq:bell_basis} similar to the teleportation procedure.

From eq~\ref{eq:swap_expanded}, a simultaneous bell state measurement at repeater station $R_x$ that is, on qubits $R_{x_1}$ and $R_{x_2}$ would result in either of four Bell states that is, $|\phi^+ \rangle^{R_{x_1} R_{x_2}}, |\phi^- \rangle^{R_{x_1} R_{x_2}}, |\psi^+ \rangle^{R_{x_1} R_{x_2}}$, and $|\psi^- \rangle^{R_{x_1} R_{x_2}}$. Meanwhile, the collective state of qubits of Alice and Bob that is, $A$ and $B$ would be in the state: $ \frac{1}{\sqrt{2}}\left( \alpha |00 \rangle + \beta |11 \rangle \right)^{AB}, \frac{1}{\sqrt{2}}\left( \alpha |00 \rangle - \beta |11 \rangle \right)^{AB}, \frac{1}{\sqrt{2}}\left( \alpha |01 \rangle + \beta |10 \rangle \right)^{AB}$, and $\frac{1}{\sqrt{2}}\left( \alpha |01 \rangle - \beta |10 \rangle \right)^{AB}$. The measurement results obtained at repeater station $R_x$ are communicated to Alice or Bob via a classical channel.
Depending upon these measurement results a single-qubit gate that is, $I$ for $| \phi^+ \rangle$, $\sigma_z$ for $|\phi^- \rangle$, $\sigma_x$ for $| \psi^+ \rangle$, and $\sigma_x \sigma_z$ for $|\psi^- \rangle$ which gets the collective state of Alice and Bob qubit transformed to an EPR pair $| \phi^+ \rangle$. This concludes the entanglement-swapping procedure as summarized in table~\ref{tab:bsa_swap}.

\begin{table}[h!]
\centering
\caption{\textbf{Bell state analysis for Entanglement swapping:} Bell state measurement at quantum repeater's ($R_x$) qubits $R_{x_1}$ and $R_{x_2}$ \textit{\textbf{(column: $R_{x_1}R_{x_2}$ results)}} collapses (projects) the combined Alice's qubit $A$ and Bob's qubit $B$ to several different bell states \textit{\textbf{(column: $AB$ result)}}. Hence, a single-qubit gate(s) is applied at Bob's qubit $B$ \textit{\textbf{(column: applied single-qubit gate at Bob (B))}} to arrive at correct end-to-end entanglement state between Alice's qubit $A$ and Bob's qubit $B$ \textit{\textbf{(column: final state ($AB$))}}.}
\label{tab:bsa_swap}
\begin{tabular}{|l|l|l|l|}
\hline
$R_{x_1} R_{x_2}$ results & $AB$ result & applied single-qubit gate at Bob ($B$) & final state ($AB$) \\
\hline
$|\phi^+ \rangle^{R_{x_1} R_{x_2}}$ & $\frac{1}{\sqrt{2}}\left( \alpha |00 \rangle + \beta |11 \rangle \right)^{AB}$ & $I$ & $\frac{1}{\sqrt{2}}\left( \alpha |00 \rangle + \beta |11 \rangle \right)^{AB}$ \\
$|\phi^- \rangle^{R_{x_1} R_{x_2}}$ & $\frac{1}{\sqrt{2}}\left( \alpha |00 \rangle - \beta |11 \rangle \right)^{AB}$ & $\sigma_z$ & $\frac{1}{\sqrt{2}}\left( \alpha |00 \rangle + \beta |11 \rangle \right)^{AB}$ \\
$|\psi^+ \rangle^{R_{x_1} R_{x_2}}$ & $\frac{1}{\sqrt{2}}\left( \alpha |01 \rangle + \beta |10 \rangle \right)^{AB}$ & $\sigma_x$ & $\frac{1}{\sqrt{2}}\left( \alpha |00 \rangle + \beta |11 \rangle \right)^{AB}$ \\
$|\psi^- \rangle^{R_{x_1} R_{x_2}}$ & $\frac{1}{\sqrt{2}}\left( \alpha |01 \rangle - \beta |10 \rangle \right)^{AB}$ & $\sigma_x \sigma_z$ & $\frac{1}{\sqrt{2}}\left( \alpha |00 \rangle + \beta |11 \rangle \right)^{AB}$ \\
\hline
\end{tabular}
\end{table}

Entanglement swapping is the basic concept implemented by quantum repeaters, introduced below, which are the fundamental building blocks of the quantum internet.

\vspace*{0.5cm}

\begin{figure}
\centering
\includegraphics[width=\columnwidth]{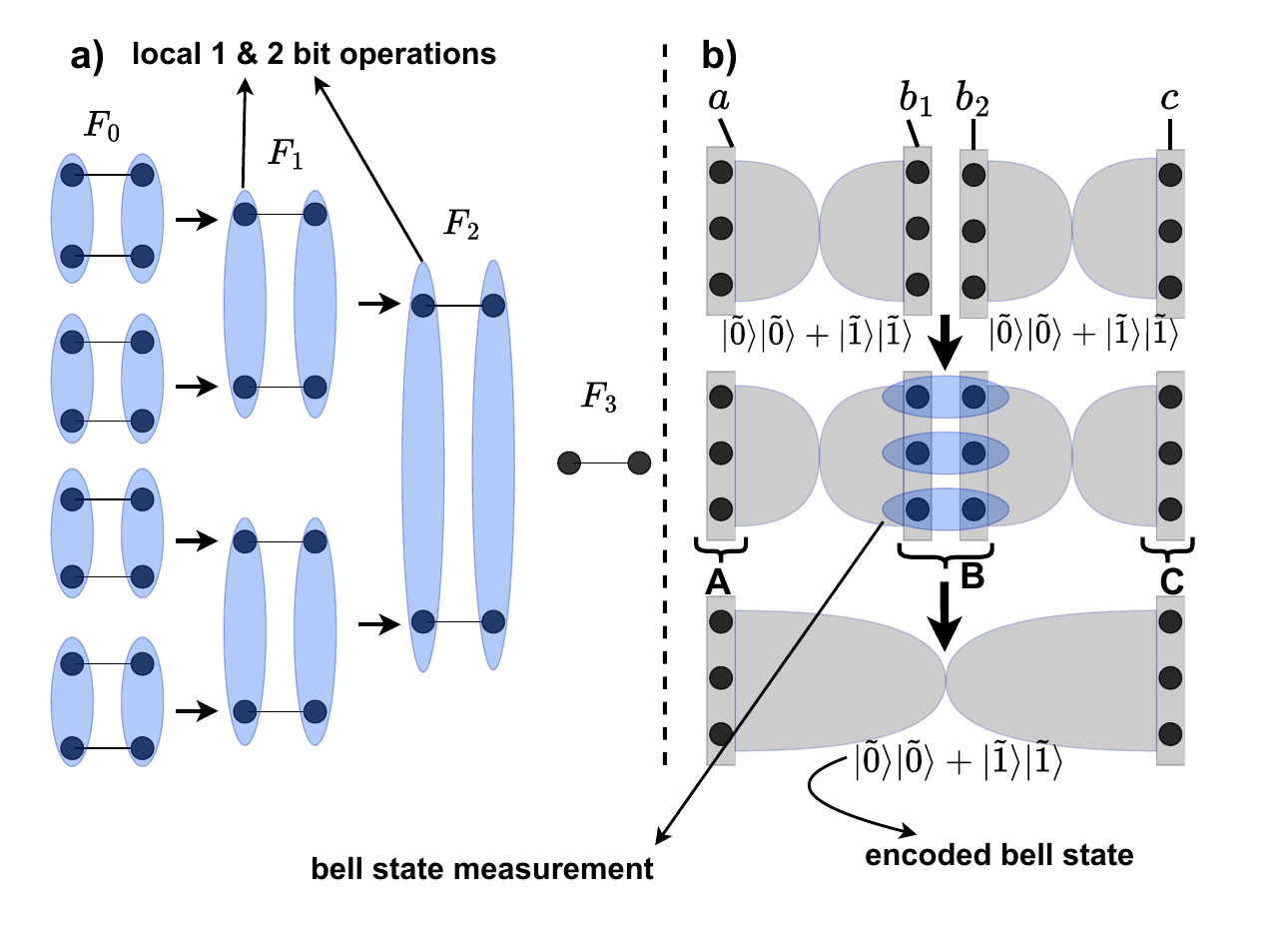}
\caption{
Using extra qubit resources to boost error resilience. \textbf{a) Entanglement purification:} Each EPR pair is paired with an additional, \textit{sacrificial} EPR pair to enhance its fidelity (modified from \cite{dur99}); \textbf{b) Quantum error correction:} Each EPR pair is supported by extra qubits arranged as \textit{encoded} EPR pairs to reduce the error rate (modified from \cite{jiang09})}
    \label{fig:purification_errorcorrection}
\end{figure}

\textbf{Entanglement Purification:}
The entanglement purification procedure uses \textit{sacrificial EPR pairs} of low-fidelity to attain higher-fidelity EPR pairs or entanglement. For example, as shown in figure~\ref{fig:purification_errorcorrection}a, eight EPR pairs (fidelity $F_0$) are consumed over three layers of purification to extract a single higher fidelity EPR pair (fidelity $F_3$).
For simplicity, we assume that the 1 $\&$ 2-bit operations always result in the success of the purification protocol.

Let the initial EPR pair that is, $|\phi^+ \rangle = \frac{1}{\sqrt{2}} \left( |00 \rangle + |11 \rangle \right)$ with qubits 1 and 2 be distributed by two neighbouring nodes via a noisy quantum channel, which degrades the quality of the original \emph{pure} quantum states, hence diminishing their fidelity compared to the original Bell state and leading to a \emph{mixed}, that is, non-pure, state. The resulting mixed Bell state after exposure to this channel can be written using the following Werner's state \cite{werner89}:
\begin{equation}
    \rho = F | \phi^+ \rangle \langle \phi^+ | + \frac{1-F}{3} \left( | \phi^- \rangle \langle \phi^- | + | \psi^+ \rangle \langle \psi^+ | + | \psi^- \rangle \langle \psi^- |
    \right)
\end{equation}
This means that the probability of finding the initial EPR pair ($\rho$) shared on exposure to a noisy quantum channel intact with respect to bell state $| \phi^+ \rangle$ is $F$. Meanwhile, the probability of finding any other state is $1-F$.

To purify this EPR pair, we share another EPR pair with qubits 3 and 4 of the same bell state as before. If node A has qubits 1 and 3 while node B has qubits 2 and 4, then CNOT quantum gates with sources qubits 1 and 2 as sources and qubits 3 and 4 as targets are operated. Node A measures qubit 3, and Node B measures qubit 4, and they exchange their measured results through the classical channel (\textit{two-way signalling}).
If the measurement results match, then the EPR pair with qubits 1 and 2 is kept, or else it is discarded. Due to the measurement, the EPR pair is no longer entangled as soon as qubits 3 and 4 are measured. However, by using this \textit{sacrificial EPR pair} the fidelity of retained EPR pair is given by \cite{bennett96}:
\begin{equation}
    F_1 = \frac{F^2 + \frac{1}{9} (1-F)^2}{F^2 + \frac{2}{3}F(1-F) + \frac{5}{9}(1-F)^2}
\end{equation}
It is to be noted that $F_1 > F$ for only $F > 0.5$.

The purification protocol discussed above represents one of the initial approaches to this technology. Subsequent protocols have introduced various enhancements. A recent survey detailing advancements in entanglement purification is available in \cite{yan23}.

Entanglement purification has no equivalent in classical digital systems because the data stored or transferred are either fully correct (a 0 is a 0, a 1 is a 1) or flipped due to errors (a 0 is a 1, a 1 is a 0).

\vspace*{0.5cm}
\textbf{Quantum Error Correction:} Quantum error correction is a critical technique used to encode quantum states in a way that makes them robust against errors, analogous to Forward Error Correction (FEC) in classical systems. In classical FEC, redundancy is introduced to recover information lost due to errors during storage (e.g., on hard drives) or transmission (e.g., over noisy wireless links). To illustrate quantum error correction in the context of quantum networks, consider the example of entanglement swapping using encoded Einstein-Podolsky-Rosen (EPR) pairs.

Imagine a three-node setup: \textbf{Node A} (with a qubit array \( a \)), \textbf{Node B} (with qubit arrays \( b_1 \) and \( b_2 \)), and \textbf{Node C} (with a qubit array \( c \)), as depicted in Figure~\ref{fig:purification_errorcorrection}b. This scenario builds upon the previously discussed entanglement swapping procedure (Figure~\ref{fig:teleport_swap_chain}b) but incorporates quantum error correction by using \textit{logical} qubits instead of \textit{physical} qubits.

A \textit{logical qubit} is an abstract representation of quantum information encoded within a set of physical qubits, making it resistant to errors through redundancy. In this setup, each qubit in a standard entanglement swapping protocol is replaced by an array of \( n \) physical qubits utilizing $n$-qubit repetition code. In our case (Figure~\ref{fig:purification_errorcorrection}b), we have \( n = 3 \), which corresponds to the use of a three-qubit repetition code.

In the broader context of error correction, it is also instructive to introduce the general notation $[[n,k,d]]$, which succinctly characterizes an error-correcting code. Here, $n$ denotes the total number of physical qubits used in the encoding, $k$ represents the number of logical qubits encoded, and $d$ is the code distance---the minimum number of physical qubit errors required to cause an undetectable error. This distance determines the error-correcting capability of the code; specifically, a code with distance $d$ can detect up to $d-1$ errors and correct up to $\lfloor (d-1)/2 \rfloor$ errors. This notation mirrors the classical code notation \([n,k,d]\). In our example utilizing the three-qubit repetition code, the parameters can be expressed as $[[3,1,3]]$; a single logical qubit is redundantly encoded into three physical qubits, and the code is capable of correcting any single-qubit error.

Instead of standard EPR pairs, entanglement swapping with error correction employs \textit{encoded EPR pairs}, requiring \( n = 3 \) physical qubits for each logical qubit. For example, as shown in Figure~\ref{fig:purification_errorcorrection}b, encoded EPR pairs are established between \textbf{Node A} and \textbf{Node B} (\( a \) and \( b_1 \)) and between \textbf{Node B} and \textbf{Node C} (\( b_2 \) and \( c \)). These encoded EPR pairs are represented as:

\[
| \Tilde{\phi}^+ \rangle_{ab_1} = \frac{1}{\sqrt{2}} \left( |\Tilde{0} \rangle |\Tilde{0} \rangle + |\Tilde{1} \rangle |\Tilde{1} \rangle \right)_{ab_1}
\]
and
\[
| \Tilde{\phi}^+ \rangle_{b_2c} = \frac{1}{\sqrt{2}} \left( |\Tilde{0} \rangle |\Tilde{0} \rangle + |\Tilde{1} \rangle |\Tilde{1} \rangle \right)_{b_2c}.
\]

Here, each logical qubit is encoded using the three-qubit repetition code, where \( |\Tilde{0} \rangle = |000 \rangle \) and \( |\Tilde{1} \rangle = |111 \rangle \). Establishing a single encoded EPR pair between two nodes involves a three-step process and requires an additional ancilla qubit for each physical qubit in the encoded EPR pair, as described in \cite{gottesman99, zhou00}. This process is known as the \textit{encoded generation} of EPR pairs.

Analogous to standard entanglement swapping, the combined state of the four logical qubits can be expressed as \cite{jiang09}:

\begin{equation}\label{eq:error_correction}
\begin{split}
    |\Tilde{\phi}^+ \rangle_{ab_1} |\Tilde{\phi}^+ \rangle_{b_2c} = \frac{1}{2} \Big( 
|\Tilde{\phi}^+ \rangle_{ac} |\Tilde{+} \rangle_{b_1} |\Tilde{0} \rangle_{b_2} +  |\Tilde{\phi}^- \rangle_{ac} |\Tilde{-} \rangle_{b_1} |\Tilde{0} \rangle_{b_2} \\ 
+  |\Tilde{\psi}^+ \rangle_{ac} |\Tilde{-} \rangle_{b_1} |\Tilde{0} \rangle_{b_2} + |\Tilde{\psi}^- \rangle_{ac} |\Tilde{-} \rangle_{b_1} |\Tilde{1} \rangle_{b_2} \Big).
\end{split}
\end{equation}

This formulation enables a modified form of entanglement swapping, where Bell-state measurements are performed on logical qubits (groups of physical qubits). Specifically, the logical qubit \( b_1 \) at \textbf{Node B} is measured in the \( \{ |\Tilde{+} \rangle, |\Tilde{-} \rangle \} \) basis, while \( b_2 \) is measured in the \( \{ |\Tilde{0} \rangle, |\Tilde{1} \rangle \} \) basis. This step is referred to as the \textit{encoded connection}.

The results of these measurements yield a two-bit classical message, which is used to establish the encoded EPR pair between \textbf{Node A} and \textbf{Node C}.

Notably, these measurement results do not need to be communicated to other nodes, as is required in standard entanglement-swapping protocols. Instead, they are utilized locally at \textbf{Node B} to determine the state of the encoded qubit.

\subsection{Components of Quantum Internet}
While a fully functional quantum internet or network is not yet available, current understanding allows us to identify its key components. As the field advances, additional components may be introduced. In this discussion, we outline the essential components required for the operation of a quantum internet based on our present knowledge.

\textbf{Quantum Repeaters:} The quantum repeater\footnote{An alternative approach to entanglement distribution, utilising percolation theory, has been proposed \cite{perseguers08, das18}. However, the prevailing focus in current literature has been on the development of quantum repeaters rather than the percolation theory approach. Nonetheless, the percolation approach continues to be studied and may represent a potential area for future investigation.} is a fundamental component of the quantum internet, playing a crucial role in the entanglement swapping process, which extends entanglement over long distances, as illustrated in Figure~\ref{fig:teleport_swap_chain}c. In a linear chain of repeaters connected by shared EPR pairs, simultaneous Bell state measurements at the quantum repeaters entangle the end nodes. This is accomplished by communicating the classical results to one of the end nodes and applying the appropriate single-qubit gate at that node. It should be noted that this process is a generalised version of entanglement swapping, involving multiple repeater nodes instead of just one. 

\begin{figure}
\centering
\includegraphics[width=\columnwidth]{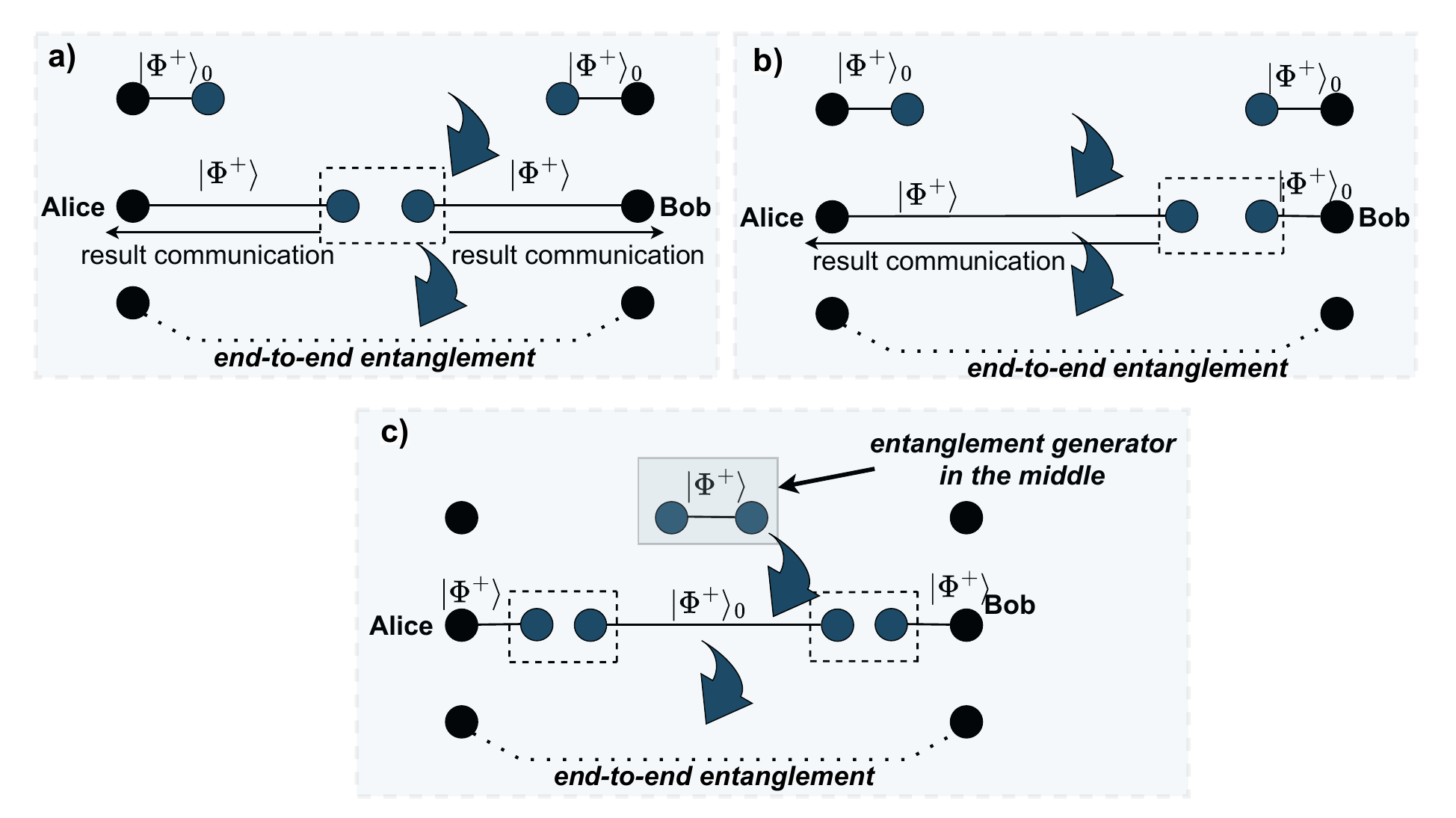}
\caption{
Entanglement generation schemes. \textit{\textbf{a) MeetInTheMiddle:}} Each of two distant stations holds one half of an EPR pair. The entangled qubits are sent to an intermediate station, where entanglement swapping establishes a direct end-to-end entanglement between the remote stations; \textit{\textbf{b) SenderRecever:}} An entangled qubit generated at the source station is transmitted to the destination station. Here, entanglement swapping between this qubit and one from the destination’s EPR pair creates a direct entanglement link between the stations; \textit{\textbf{c) MidPointSource:}} An intermediate station produces an EPR pair and sends one qubit to the source station and the other to the destination station. At each station, entanglement swapping between the received qubit and a qubit from the local EPR pair sets up end-to-end entanglement. }
    \label{fig:entanglement_generation}
\end{figure}

A quantum repeater may be equipped with a \emph{quantum memory}, that is, a device that can store quantum states as qubits for a limited amount of time, and may also include a \emph{quantum processor}, that is, equipment for the execution of local operations through quantum gates. The requirement for quantum memory in a quantum repeater depends on the intended application and the entanglement distribution protocols employed by the quantum network. For instance, in prepare-and-measure quantum key distribution (QKD) networks, quantum memories are not necessary, whereas more advanced quantum applications, such as distributed quantum computing, rely on quantum memory to function effectively.
From a protocol perspective, the need for quantum memory also varies. In networks where entanglement generation and swapping proceed sequentially or with minimal delay to directly entangle end nodes, quantum memory support may not be essential. However, if the protocol is designed such that, following initial entanglement generation, some links must wait for entanglement swapping to proceed, then quantum memory support becomes crucial. Quantum memory is also indispensable for protocols that involve purification or quantum error correction.
Similarly, the inclusion of a quantum processor depends on the network protocols. If certain protocols are implemented that leverage a quantum processor to optimize the performance of the quantum network, then it becomes necessary. In the absence of such protocols, a quantum processor may not be required. An \emph{entanglement generator} can be part of a quantum repeater if the entanglement generation scheme follows either the \textit{MeetInTheMiddle} or \textit{SenderReceiver} configuration \cite{jones16}. In the MeetInTheMiddle scheme, entanglement generation occurs at the outer nodes, with one qubit of the entangled pairs from each outer node sent to an intermediate station where entanglement swapping takes place. In contrast, in the SenderReceiver scheme, entanglement generation also occurs at the outer nodes, as in the previous case. However, one qubit of the entangled pair from one outer node is sent directly to the other outer node, eliminating the need for an intermediate station. For the \textit{MidpointSource} scheme, a separate entanglement generation component positioned at the midpoint of a quantum network link would be required, as shown in Figure~\ref{fig:entanglement_generation}.

\vspace*{0.5cm}

\textbf{Quantum Device:} A quantum device, often referred to as an \textit{end node}, is essentially a quantum computer capable of running specific quantum applications while connected to the quantum internet. These devices are equipped with quantum processors to handle the execution of quantum applications and quantum memories to manage incoming entangled quantum states.

\vspace*{0.5cm}

\begin{figure}
\centering
\includegraphics[width=\columnwidth]{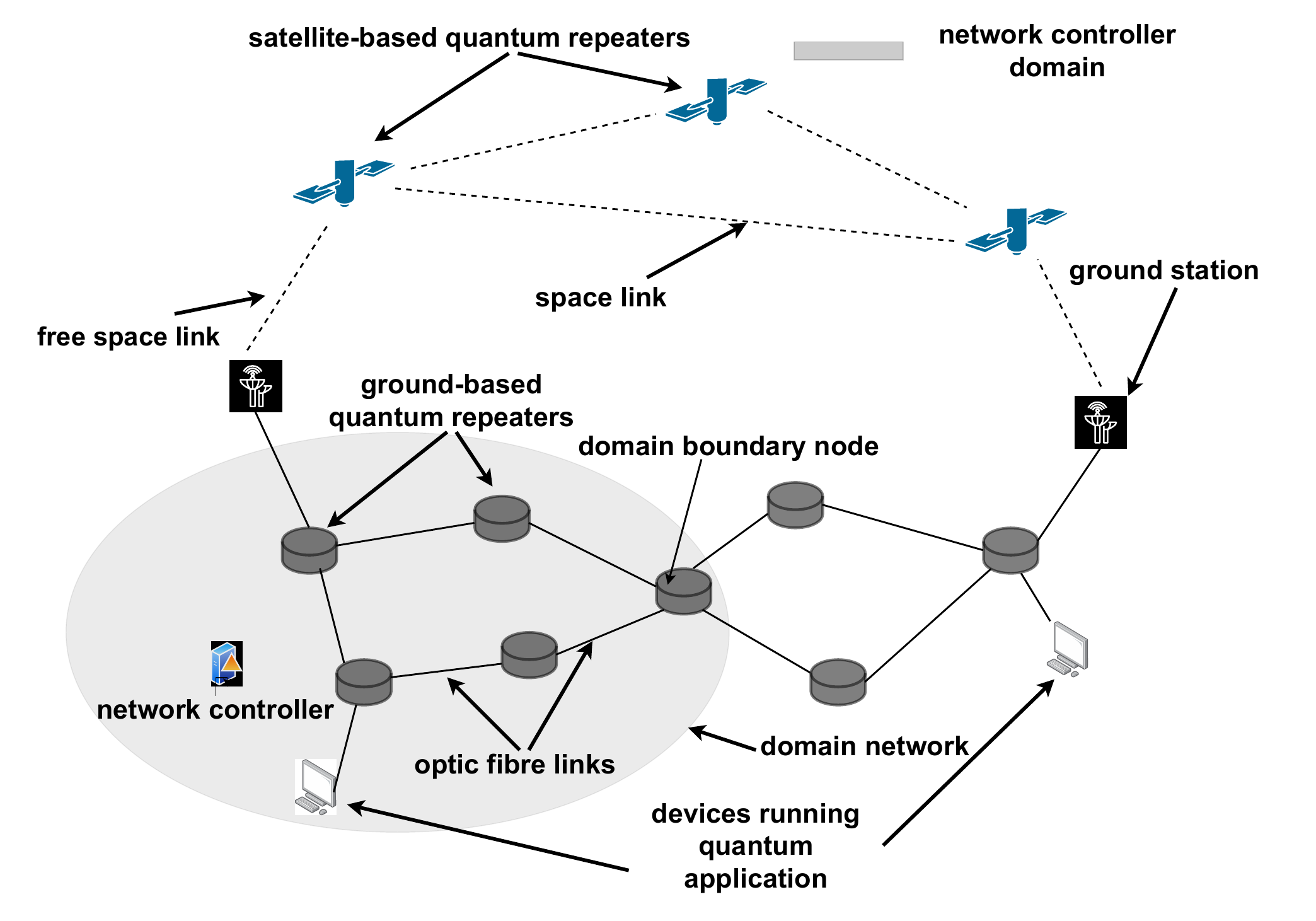}
\caption{
A schematic of the \textit{global} quantum internet.}
    \label{fig:network}
\end{figure}

\textbf{Quantum links:} The quantum links are the quantum and classical channels that connect two neighbouring quantum repeaters or a quantum repeater and a quantum device. These links can be realised through fibre optic cables (represented as solid lines) or via free-space communication (represented as dotted lines), commonly used in ground-satellite networks, as shown in Figure~\ref{fig:network}.

Because of current technology limitations, the generation of EPR pairs between neighbouring components is a probabilistic process, whose success rate depends on the specific technology used and the distance between the nodes, but in general is rather small ($\ll 0.5$).

This aspect highlights the inherent variability and complexity in the structure of quantum networks.

\vspace*{0.5cm}

\textbf{Network Controller:} A network controller is a logically centralised entity responsible for overseeing a specific segment of the quantum network within its jurisdiction. It manages the communication of Bell state measurement results from quantum repeaters to the end nodes, facilitating the completion of the entanglement swapping process. Additionally, the network controller handles routing decisions for incoming requests, ensuring efficient network operation.
The network controller is commonly found in many proposed architectures of quantum networks, even if it may be an obstacle to scalability as the network grows, both in geographical size and the number of nodes.
Therefore, decentralised alternatives are also under study to cover such use cases, even if they are currently less explored due to the additional complexity required.
Only time will tell which approach will dominate the future of the quantum internet.

\section{Applications and Use Cases}\label{sec:application}
Just as quantum computing provides advantages over classical computing, the potential applications of the quantum internet are expected to unfold as the technology advances. However, several applications have already been identified. These applications and use cases of the quantum internet can be categorised as follows.

\subsection{Enhanced security or privacy} The applications in this category essentially hinge on providing uncompromising security or privacy compared to their classical counterparts. As introduced in Section~\ref{sec:qi_fundamentals}, these applications use the fundamental principles of quantum mechanics, that is, the no-cloning theorem and quantum entanglement.

The advancements in quantum computing pose a significant threat to traditional public key cryptographic systems such as RSA \cite{rivest78}, Diffie-Hellman \cite{diffie22}, and ECC \cite{miller85, koblitz87}. To address this challenge, two primary solutions are proposed. The first involves utilising quantum cryptography, specifically \textbf{quantum key distribution} (QKD) \cite{nurhadi18, mehic20, cao22}, which is one of its most successful applications. QKD protocols allow two parties—such as quantum devices within the quantum internet—to share symmetric secret keys securely. These keys can then be used with encryption algorithms to transmit messages securely. The second solution is the adoption of post-quantum cryptography \cite{bernstein17}, which employs classical systems designed to be secure against quantum attacks.

Once a quantum device connects to the quantum internet, it can utilise this network much like the classical internet we use today. In this environment, the device can offload quantum computation tasks to an untrusted device without risking the privacy or integrity of its data. This security method is termed \textbf{blind quantum computing} (BQC) \cite{broadbent09, fitzsimons17}. Essentially, a client quantum device can employ one or more servers to perform computational tasks while concealing the nature of the computations from the servers themselves. Furthermore, some BQC protocols enhance security by incorporating the ability to verify the computations performed by the servers. This is achieved by embedding hidden tests within the computations, ensuring their correctness and integrity.

Following the theme of enhancing security and privacy, quantum technology offers various applications in distributed systems related to \textbf{consensus and verification}. A prime example is the quantum Byzantine agreement, developed by Ben-Or et al. \cite{ben05}, which ensures that a group of participants can reach consensus on a bit value, robust against faulty or malicious behaviors. Building on this foundation of trust, quantum technology extends its utility to secure voting and surveying systems. As outlined by Vaccaro et al. \cite{vaccaro07}, these systems allow participants to cast votes or respond to surveys with guaranteed anonymity, enhancing the integrity of collective decision-making processes. Moreover, the field of quantum cryptography has made significant strides in enabling secure multiparty communications. Quantum Conference Key Agreement (CKA), as discussed in recent studies by Hahn et al. \cite{hahn20} and Murta et al. \cite{murta20}, facilitates the establishment of a shared secret key among multiple parties, crucial for coordinating actions across different nodes without compromising security. Another innovative application is certified deletion, introduced by Broadbent et al. \cite{broadbent20}, which allows for the verifiable destruction of encrypted information, ensuring that once a ciphertext is deleted, it remains irrecoverable, even if the decryption key is compromised. Further expanding the horizon, quantum leader election protocols, such as those explored by Ganz et al. \cite{ganz17}, enable a group of distant, mutually distrustful entities to democratically elect a leader, ensuring fairness and transparency in critical decision-making scenarios. Additionally, the integration of quantum technologies into broader infrastructures is underway, with applications in the Internet of Things (IoT) being actively researched by Rahman et al. \cite{rahman19} and prospective uses in future 6G networks as envisioned by Rozenman et al. \cite{rozenman23}. These advancements illustrate the expanding role of quantum technology in shaping modern technological landscapes, promising unprecedented levels of security and efficiency in digital communications and beyond.

\subsection{Enhanced computing}
Applications within this category significantly enhance computing capabilities, enabling the completion of tasks that are challenging for classical computers or even a single quantum computer to perform. This augmentation is crucial for solving complex problems that exceed the processing power of conventional computing paradigms.

The leading concept in this category is \textbf{distributed quantum computing} \cite{cacciapuoti19, caleffi24}. Distributed quantum computing involves leveraging the computational resources of multiple quantum devices interconnected via the quantum internet to perform complex computational tasks. This approach is vital given the current presence of noise in practically realised qubits and the inherent limitations on the number of qubits per computer. In the absence of fault-tolerant qubits, distributed quantum computing offers a viable solution to scale up the number of qubits, enabling the execution of tasks that are too complex for single devices. Distributed quantum computing promises exponential speed-ups as compared to linear scaling in distributed classical computing \cite{cuomo20}.

Another application of great potential interest is \textbf{quantum federated learning}\cite{sunkel24}: each client trains their model on their local dataset using a quantum computer, thereby keeping their data private. Instead of sharing private data, clients only exchange their model's weights. The global model is then trained by aggregating these weights from all participating clients. While this is done today by exchanging only classical data, there are huge opportunities that can be unlocked by entanglement via the quantum internet since this would remove the need for the data to be decoded and re-encoded at each iteration.
Finally, we mention that solutions have been proposed to enhance the accessibility and efficiency of quantum computing resources for diverse applications. Currently, they require only classical communications. For instance, \textbf{Quantum-as-a-Service}~\cite{garcia21, moguel22} involves a Quantum API Gateway that recommends the most suitable quantum computer for running a specific quantum service in real time. We can speculate that, as the quantum internet is deployed, such systems will evolve into more sophisticated versions building on the end-to-end distribution of entanglement among remote nodes.

\subsection{Specialised applications}

In addition to the general applications of the quantum internet, several specialised applications also play a crucial role in leveraging quantum mechanical capabilities, especially for scientific experiments.

One such specialised application is \textbf{quantum sensing} \cite{degen17}, which utilises the ability to establish quantum entanglement across networks to enhance the precision of measurements beyond the classical limits. This application exploits quantum properties to achieve superior measurement accuracy in various scientific and technological fields.

Another important application is \textbf{time synchronisation} in digital clocks \cite{chuang20}. Unlike classical methods that require $\bigO (2^{2n})$ messages to determine the $n$ digits of time difference $\Delta$ between two separated clocks in space, quantum algorithms can achieve the same with only $\bigO (n)$ quantum messages. This reduction in message complexity makes quantum time synchronisation significantly more efficient.

Additionally, there is the application of \textbf{enhanced-baseline length for telescopes} \cite{gottesman12, czupryniak23}. Traditional optical interferometers are limited in their resolution due to restricted baseline lengths, noise, and signal loss during photon transmission between telescopes. The use of quantum internet can potentially overcome these limitations, allowing for interferometers with arbitrarily long baselines and thus dramatically improving their resolving power.

\begin{table}[h!]
\centering
\caption{Stages of quantum internet development}
\label{tab:qi_stages}
\begin{tabular}{p{0.6cm}p{3.9cm}p{7.5cm}}
\toprule
Stage & Network & Functionality  \\
\midrule
\textit{S-1}  & QKD networks \newline
& Supports basic quantum key distribution between any nodes using trusted repeaters and post-selected prepare-and-measure techniques without end-to-end entanglement. \\

\textit{S-2} & Entanglement distribution networks & end-to-end entanglement between any nodes with no quantum memories \\
\textit{S-3} & Quantum memory networks &  end-to-end entanglement between any nodes with the capability of storing in quantum memories \\
\textit{S-4} & Limited-qubits fault-tolerant networks & end-to-end entanglement between any nodes with limited-qubits fault-tolerant capability on the quantum memory qubits \\
\textit{S-5} & Quantum computing networks & end-to-end entanglement between any nodes with full-fledged capability of using qubits in the quantum memory for computation and communication  \\
\bottomrule
\end{tabular}
\end{table}

\section{The Quantum Internet}\label{sec:architecture}

We now delve into the main topic of this work.
In Section~\ref{ssec:stages}, we introduce a possible roadmap of the quantum internet.
Then we proceed in a bottom-up manner by illustrating the state-of-the-art physical connectivity technologies available (Section~\ref{ssec:physical_layer}), the quantum repeater (Section~\ref{ssec:repeater}), the main quantum network algorithms and protocols (Section~\ref{ssec:routing_forwarding_scheduling}), and finally we provide an overview of the stack models of the quantum internet that have been proposed so far, in Section~\ref{ssec:protocol_stack}.

\subsection{Stages of quantum internet}\label{ssec:stages}
As with any developing technology, especially as sophisticated as the quantum internet, the implementation is only possible in stages. The different stages of the quantum internet have been categorised by the amount of incremental functionality

available to the quantum devices or end nodes in~\cite{wehner18}. The summary of the stages of quantum internet development is given in Table~\ref{tab:qi_stages}. 

Stage 1 of the quantum internet initiates the quantum internet with basic point-to-point quantum key distribution (QKD) setups that depend on trusted intermediate nodes for key relay and security \cite{scarani09, salvail10}. It primarily features networks where end-to-end quantum communication is absent, relying on secure key generation between adjacent nodes using trusted nodes. To bolster security against potentially untrusted nodes, measurement-device-independent QKD protocols are employed, enhancing security without relying on the trustworthiness of measurement devices \cite{lo12}. At this stage, the lack of end-to-end qubit entanglement prevents support for distributed quantum computing or quantum sensing. However, progress toward these goals begins with the ability of nodes to prepare and transmit a one-qubit state to any other node in the network. The transmission and measurement processes leverage post-selection, wherein only successful events—those where qubits are detected and measured correctly—are retained. Undetected or ``lost" qubits are disregarded.

This post-selected distribution of entanglement works as follows: a sender node prepares a pair of entangled qubits, retains one, and sends the other to a receiving node. If the receiving node successfully detects the transmitted qubit, the entanglement is confirmed and preserved. Although this method does not enable the deterministic transmission of arbitrary quantum states, it establishes a foundation for more sophisticated quantum operations in future stages.

Stage 2 of the quantum internet enables end-to-end entanglement without the need for post-selection during transmission or measurement, as in the case of Stage 1 above. However, due to the absence of quantum memory in the network, the entanglement must be used immediately after its creation. This stage allows for the successful distribution of end-to-end entanglement with a probability approaching 1. 

Stage 3 quantum internet upgrades to having the support of quantum memories at the local nodes in the network for application purposes. The quantum memories at network nodes allow more complex operations such as entanglement purification, quantum error correction and the creation of multi-partite states from bi-partite entanglement. However, due to the limited decoherence capabilities of the quantum memories, fault tolerance remains an issue in such networks. A functioning quantum memory network should have a decoherence time which encompasses the phase of entanglement generation and the time it takes for the classical signal to complete the entanglement distribution. This stage also provides the capability of deterministically sending arbitrary quantum states from one node to another.

Stage 4 of the quantum internet introduces fault-tolerance capabilities for quantum memory qubits, though these capabilities are limited to a finite number of qubits. Fault tolerance refers to the suppression of errors through the use of additional resources, namely, increased quantum memory. A group of error-corrected physical qubits is referred to as a \textit{logical qubit}. However, the suppression of errors through the combination of physical qubits is only feasible if the physical error rate remains below a critical threshold. Recent advancements in this area have demonstrated that surface code memories, even when operating below this threshold, can effectively suppress the logical error rate \cite{acharya24}.

Stage 5 of the quantum internet represents the realization of a fully developed quantum internet, enabling the complete range of applications in both computation and communication. At this stage, fault tolerance is achieved for all available quantum memory qubits, unlocking the full potential of the quantum internet for diverse and robust applications.

The roadmap above was published in 2018~\cite{wehner18}, but at the time of writing it can be considered still valid and useful to identify the upcoming milestones.
Today, stage~1 operational networks have been deployed in semi-commercial environments, as discussed later in Section~\ref{sec:current_initiatives}, while all other stages appear more elusive, in particular, due to the lack of commercial-grade products for components such as the quantum repeater (see Section~\ref{ssec:tech_challenges}).
However, significant progress has been demonstrated for many enabling technologies and quantum networks that can be categorised as stage~2 have been implemented in controlled environments, which gives us hope that technology will soon catch up with high expectations from the scientific community.

\subsection{Types of quantum network}\label{ssec:physical_layer}

According to the current consensus in the field, the quantum internet is envisioned to operate alongside the classical internet, particularly during its initial phases \cite{garcia24}. Therefore, the categorisation of quantum networks that collectively constitute the quantum internet parallels that of the classical internet, based on the operational area's size. Broadly, quantum networks can be categorised into \textit{modular} networks, \textit{ground-based} networks, and \textit{satellite-based} networks. 

\textit{Modular networks} are distinct in their construction, often associated with a multi-core quantum architecture. They are predominantly used in distributed quantum computing \cite{caleffi24} and are typically implemented on a chip to interconnect various quantum computing modules \cite{rodrigo21, jnane22}. These networks are crucial for the scalability of quantum computing technologies, providing essential links within and between quantum processors.

On the other hand, \textit{ground-based networks} can be further subdivided according to their operational distances, similar to classical networks. The most frequently discussed type of ground-based networks in current research—and those that often have experimental test beds—are metropolitan quantum networks. Metropolitan quantum networks operate over metro-scale distances \cite{chen21, chung22}, facilitating regional connectivity within a more confined geographic area compared to their long-range counterparts.

The motivation for including \textit{satellite-based networks} in the global internet stems from the fact that free space links suffer polynomial loss compared to an exponential loss in optical fibre links \cite{wallnofer22}. Therefore, satellite-based quantum networks unlock quantum communication over continental and intercontinental ranges, which is otherwise challenging with purely ground-based quantum networks. However, the free space link (Forward $\&$ Return link) passes through the atmosphere, which induces several effects, such as diffraction, absorption, and scintillation effects \cite{chiti24}. While research efforts are required on this front, the perseverance of entanglement in good weather conditions can be seen as a potential to include satellite-based networks in the quest for the quantum internet \cite{armengol08}. Given the vast distance of operation in satellite-based networks, latency also plays a huge part when considering the decoherence of quantum states. Quantum memories would play a vital role in such scenarios. Studies are ongoing on this front for the use of quantum memories in space \cite{gundogan21}. 

\subsection{Generations of quantum repeater}\label{ssec:repeater}
Quantum repeaters, as introduced in Section~\ref{sec:qi_fundamentals}, are essential for the distribution of end-to-end entanglement, especially in a long-distance regime. However, two major challenges complicate this task: \textit{photon loss} and \textit{operational errors}. Photon loss occurs when photons, the carriers of quantum information, are absorbed or scattered during transmission, preventing successful entanglement. Operational errors, on the other hand, stem from imperfections in the devices that manipulate and measure quantum states, potentially leading to incorrect results.

To address these issues, quantum repeaters are classified into three generations based on the methods they employ to correct and mitigate errors \cite{munro15, muralidharan16}.

\begin{figure}
\centering
\includegraphics[width=\columnwidth]{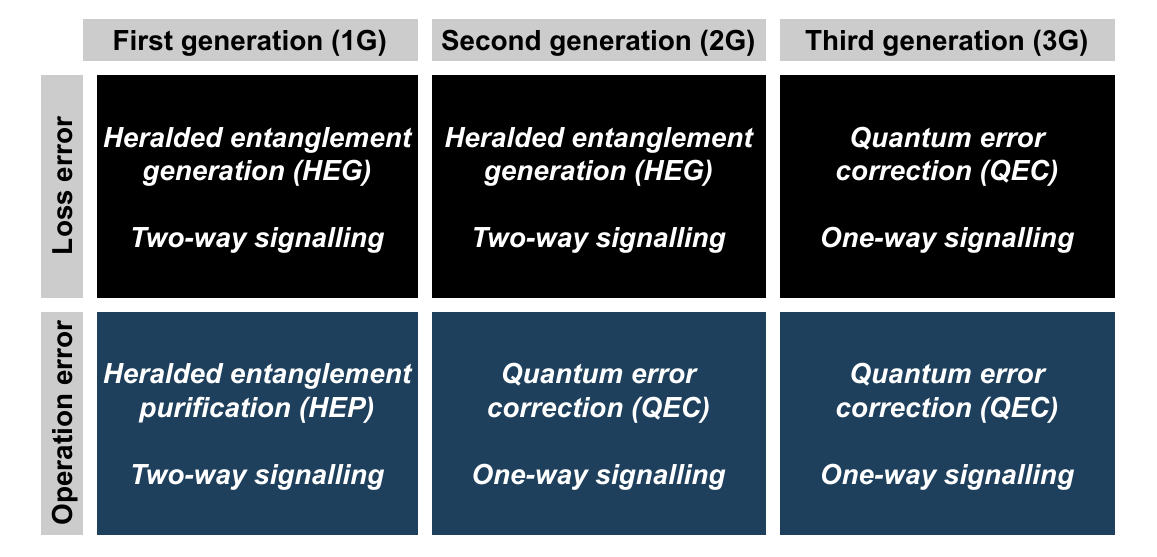}
\caption{Generations of quantum repeaters based on the method used to correct the error (modified from \cite{muralidharan16}).}
    \label{fig:repeater}
\end{figure}

The first generation of quantum repeaters employs two key techniques: \textit{heralded entanglement generation} (HEG) and \textit{heralded entanglement purification} (HEP). HEG is responsible for correcting loss errors through a deterministic process that ensures the reliable delivery of entanglement. It utilizes two-way signalling to confirm the successful establishment of end-to-end entanglement. Similarly, HEP addresses operational errors by using two-way signalling to guarantee the effectiveness of the entanglement purification process. This technique has been further elaborated in Section~\ref{sec:qi_fundamentals}. 

The second generation of quantum repeaters uses HEG to correct loss errors while \textit{quantum error correction} (QEC) to correct operational errors. HEG in the second generation of quantum repeaters still uses two-way signalling, while QEC only requires one-way signalling, which reduces the time consumption in the establishment of the end-to-end entanglement.

The third generation of quantum repeaters uses QEC to correct loss and operational errors. Subsequently, only one-way signalling is enough for the operation of the third generation of quantum repeaters, which greatly reduces the time consumption in the end-to-end entanglement establishment procedure. 

Today, the industry and academy are working towards realising stable 1G quantum repeaters, which are not yet ready for mass production and field deployment. For this reason, many of the scientific papers published, including those cited in this paper, focus on 1G repeaters only, while 2G/3G repeaters are currently mainly a matter of long-term speculation.

\subsection{Routing, forwarding, and scheduling}\label{ssec:routing_forwarding_scheduling}
As introduced in Section~\ref{ssec:basics}, the applications in Section~\ref{sec:application} utilise two processes. First, transporting an arbitrary quantum state using a quantum teleportation protocol. Second, swapping entanglement is distributed over smaller distances with an end-to-end entanglement between two nodes or devices using an entanglement-swapping protocol. While quantum teleportation is necessary for quantum communication, establishing end-to-end entanglement unlocks most applications. It is to be noted that the two protocols are almost similar and inter-convertible with only the difference of an extra qubit and bell state measurement involved as depicted in Figure~\ref{fig:teleport_swap_chain}.

\begin{figure}
\centering
\includegraphics[width=\columnwidth]{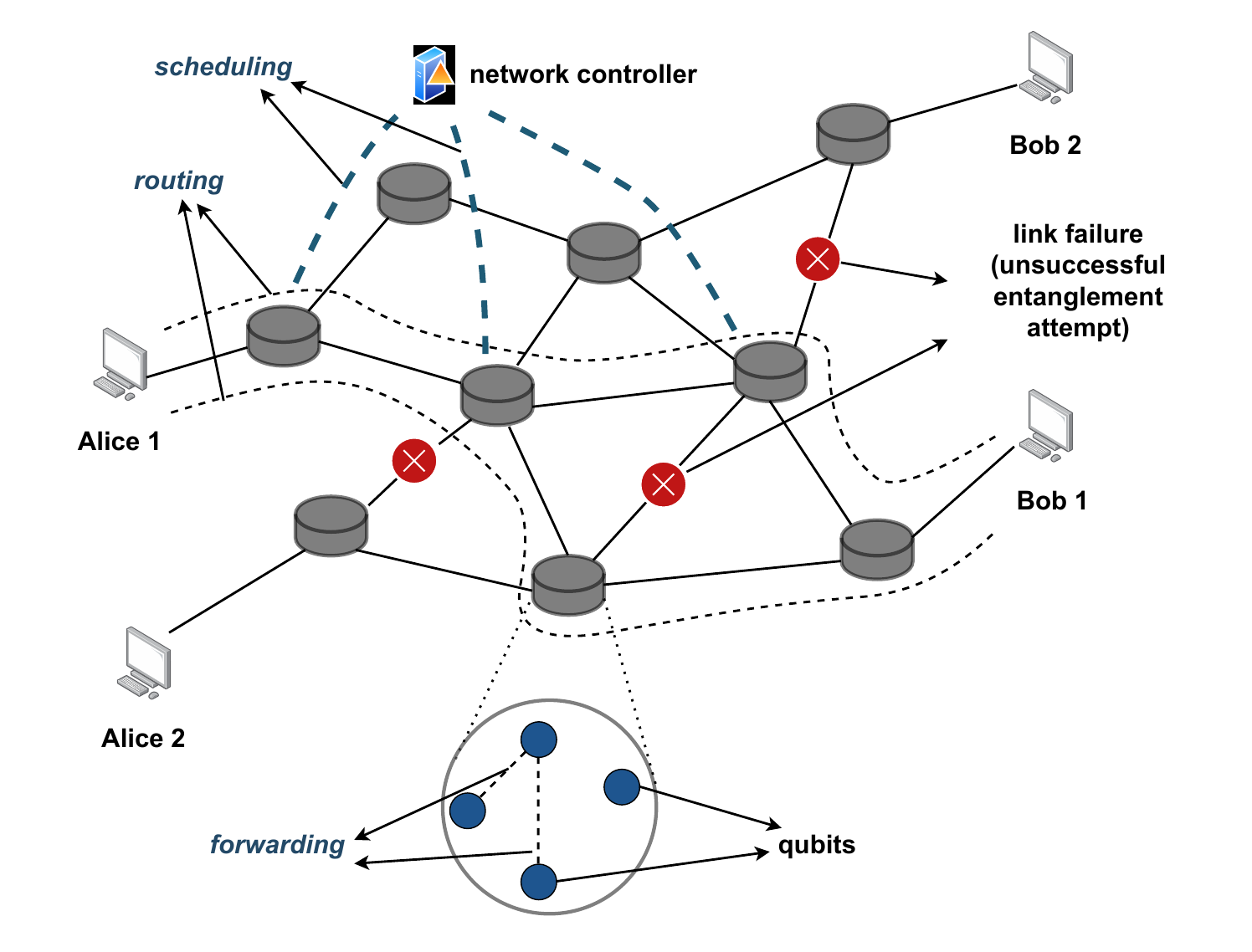}
\caption{\textbf{Routing:} selection of route. \textbf{Forwarding:} execution of entanglement swapping and purification procedure. \textbf{Scheduling:} selection of end-to-end entanglement in a time-slot.}
    \label{fig:routing}
\end{figure}

Achieving end-to-end entanglement between two quantum devices within a network that includes quantum repeater nodes is a complex challenge under practical constraints. Let us delve into the intricacies of this problem. Imagine a network of quantum repeaters connected by quantum links, as depicted in figure~\ref{fig:routing}. The network controller receives requests to establish end-to-end entanglement between the connected quantum devices. The network operates using specific protocols for entanglement generation \cite{caleffi17}, entanglement purification \cite{victora20, li22, hu24}, and quantum error correction \cite{patil24}. Additionally, the components of the network, including quantum links \cite{meter13}, repeaters \cite{vk23, vk24}, and devices, could be heterogeneous; that is, they vary either in quality or involve different physical systems.

Given these protocols and assumptions, the \textit{problem} involves determining how to \textit{satisfy} the requests received by the network controller\footnote{For simplicity in explanation, we assume the presence of a network controller, as this is a common approach in the literature. However, it is worth noting that approaches not requiring a network controller are also feasible as discussed in Section~\ref{ssec:simulation}.} in a \textit{practical scenario} by selecting the most efficient routes, swapping orders, and scheduling policy within the network to effectively serve the requests referred to as \textit{routing}, \textit{forwarding}, and \textit{scheduling} respectively. To \textit{satisfy} means to serve these requests with the highest possible throughput \cite{zhao21} and fidelity—meeting or exceeding the threshold fidelity set by the requests—while minimising latency, ensuring fairness, and optimising resource \cite{zhang21}. The \textit{practical scenario} means taking into account operational challenges such as signal attenuation due to distance, noise during the application of quantum gates in the procedure, quantum state decoherence over time, and depolarisation due to the quantum channel. 

For multiple end-to-end entanglement deliveries between source-destination pairs in a quantum network, finding an optimal path using some routing metric (discussed in Section~\ref{ssec:simulation}) is referred to as \textit{routing}. 

\begin{figure}
\centering
\includegraphics[width=0.8\columnwidth]{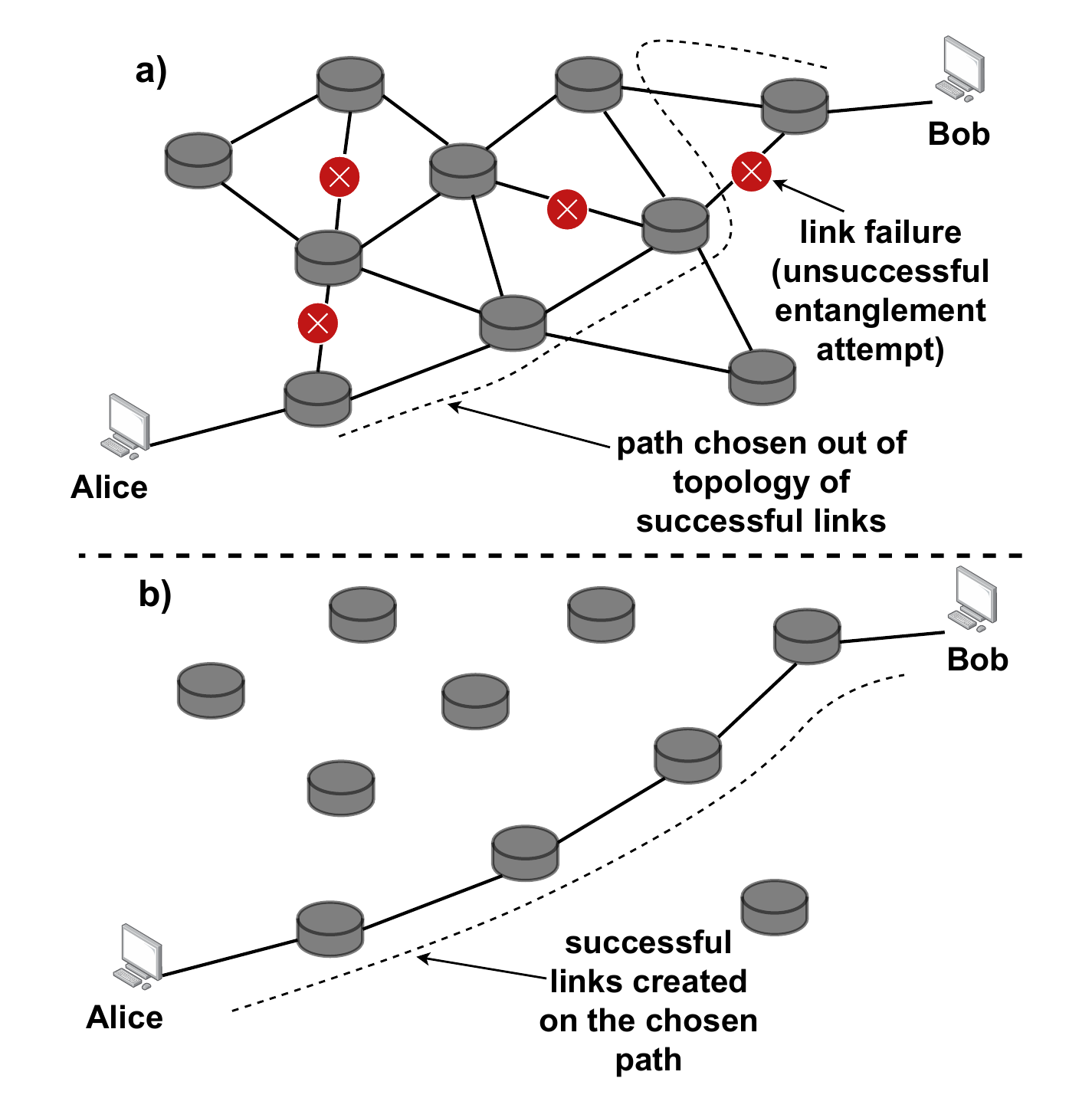}
\caption{\textbf{a) Advanced entanglement generation:} The Path is chosen out of a reduced topology constructed out of successful entanglement links. \textbf{b) On-demand entanglement generation:} Successful entanglement links are established on a chosen path.}
    \label{fig:advanced_and_on_demand}
\end{figure}

Meanwhile, the actual execution of the entanglement swapping procedure and purification (depending upon the protocol), which considers the path found using the routing algorithm, is referred to as \textit{forwarding}. The routing phase can further be categorised based on entanglement generation utilised, that is, on-demand or advanced entanglement generation (depicted in figure~\ref{fig:advanced_and_on_demand}). On-demand entanglement generation calculates routing paths before initiating entanglement, whereas advanced entanglement generation bases its routes on the network topology that emerges following the probabilistic success or failure of initial link establishment. Each approach offers distinct benefits and limitations. On-demand generation allows routing decisions to be made before quantum states begin to decohere, which is advantageous. However, due to the inherently probabilistic nature of entanglement, it may require multiple attempts to secure all necessary links for a successful route. On the other hand, advanced entanglement generation relies on post-link-establishment topology data. While this method integrates more recent network information, it risks the routing decisions occurring within the decoherence window of the entangled photon pairs, which could degrade the quality of the entanglement. The choice of entanglement generation scheme significantly impacts the quantum network's requirements. For on-demand entanglement generation, a study \cite{khatri19} has proposed two figures of merit. The first is the average connection time, which dictates the quantum memory requirements. The second is the average largest entanglement cluster size, which indicates the scalability of the quantum networks. In the literature, the routing phases used in on-demand and advanced entanglement generation are also referred to as proactive and reactive, respectively. 

The issue of \textit{scheduling} has recently begun to attract attention in the field of quantum networking. Scheduling involves selecting which end-to-end entangled EPR pairs to allocate to a current path request within a specific time-slot (discussed in Section~\ref{ssec:simulation}) while reserving EPR pairs for future path requests \cite{cicconetti21, wang23, fittipaldi22, gu23}. This process is crucial for managing latency, particularly due to the decoherence of quantum states. Effective scheduling ensures that the quantum states are utilised efficiently, minimising the impact of decoherence on network performance and maximising the fidelity of quantum communications.

It is evident that addressing all potential constraints to deliver entanglement in a multi-user quantum network presents substantial difficulties. For instance, optimising end-to-end entanglement for multiple source-destination pairs is an NP-hard problem, as identified in \cite{chakraborty20}. Additionally, the specific type of application running on the network significantly influences the demands placed on it, which suggests that distinct routing problems could be formulated for each application class.

The routing problem described earlier, which focuses on delivering end-to-end entanglement between specified endpoints, is particularly relevant for applications that enhance security or privacy, as well as for other specialised applications discussed in Section~\ref{sec:application}. Conversely, application classes that primarily enhance computing capabilities, such as those found in distributed quantum computing, have distinct requirements that necessitate alternative routing solutions \cite{cicconetti22, cicconetti23}. These variations underscore the complexity and specialised nature of routing in quantum networks.

\begin{table}[h!]
\centering
\caption{A general recipe for end-to-end entanglement distribution in a quantum network}
\label{tab:recipe_entanglement}
\begin{tabular}{p{3.6cm}p{8.5cm}}
\toprule
Stage & Choices  \\
\midrule
Ingredients & Bell (or GHZ) states  \\

Routing Algorithms & Dijkstra-based, optimisation programs (linear, integer, stochastic, etc), greedy-based, AI-based, etc.  \\
Evaluation metrics & Throughput, fidelity, hop count, etc. \\

Performance enhancers & Purification, error correction  \\

\bottomrule
\end{tabular}
\end{table}

\subsection{Quantum internet protocol stack}\label{ssec:protocol_stack}
\begin{figure}
\centering
\includegraphics[width=\columnwidth]{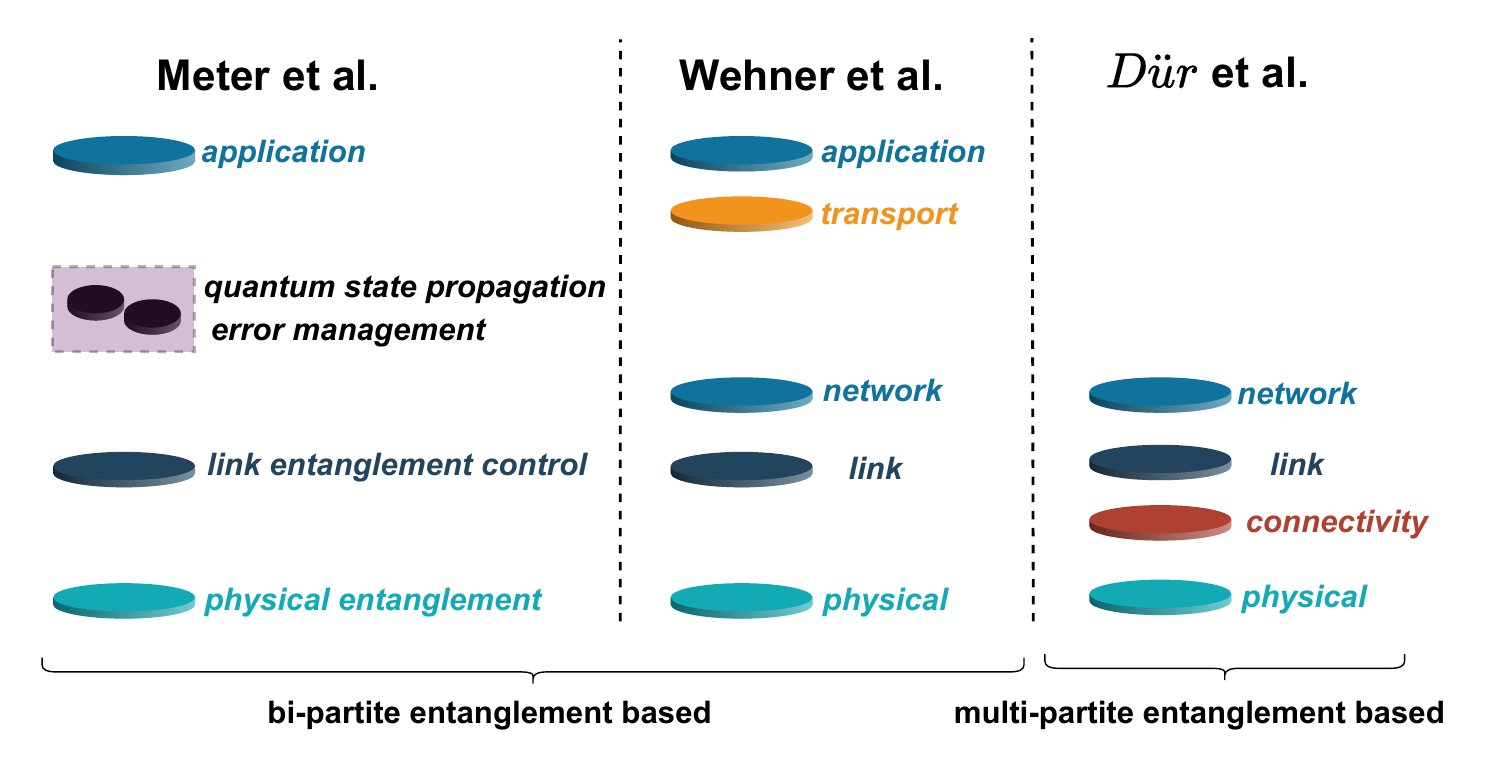}
\caption{State-of-the-art quantum internet protocol stacks. Meter et al. \cite{aparicio11}, Wehner et al. \cite{kozlowski20}, and D\"{u}r et al. \cite{pirker19}.}
    \label{fig:stack}
\end{figure}

For efficient and scalable network development, a protocol stack is indispensable, as it offers a structured framework enabling the independent development of each layer. Numerous research groups have put forth proposals for a quantum internet protocol stack, similar to the classical internet’s Open Systems Interconnection (OSI) model, which segregates network functions into distinct layers. In this paper, we will examine three quantum internet protocol stacks that are most frequently referenced in the scholarly literature. In Figure~\ref{fig:stack}, we provide a graphical visualisation of the stacks, highlighting their differences. For more details on this specific topic, interested readers are invited to check recent relevant surveys, like~\cite{illiano22, li24}.

The initial comprehensive proposal for a layered quantum internet protocol stack dates back to 2009, introduced by Meter et al. \cite{meter08, aparicio11}. This protocol stack comprises five layers: \textit{physical entanglement}, \textit{link entanglement control}, \textit{error management}, \textit{quantum state propagation}, and \textit{application}. The \textit{physical entanglement} layer is tasked with generating EPR pairs between adjacent nodes, while the \textit{link entanglement control} layer monitors the success and failure of these pair establishment attempts. These two layers operate across every single hop and are applied recursively to maintain connectivity through shared EPR pairs. The \textit{error management} layer specifies the hops where purification is needed, which ensures the high fidelity of the EPR pairs and records the outcomes. The \textit{quantum state propagation} layer is responsible for establishing end-to-end entanglement by creating shared EPR pairs between the source and destination stations using the entanglement swapping procedure. This layer also communicates the results of the swapping to the end nodes, enabling them to perform the necessary single-qubit operations. Both the \textit{error management} and \textit{quantum state propagation} layers are similarly operated recursively across multiple hops to ensure high fidelity of the distributed entanglement. Finally, the \textit{application} layer primarily manages the applications run on the quantum internet using high-fidelity EPR pairs. Recent advancements in this protocol stack have focused on ensuring synchronicity between distant nodes \cite{matsuo19}. This study introduced a RuleSet-based quantum link bootstrapping protocol that assesses the fidelity of quantum links and their throughput. Another study implemented a quantum recursive network architecture (QRNA) alongside the RuleSet-based protocol to enhance scalability, specifically achieving multi-party entanglement and internetworking within quantum networks \cite{meter22}.

The protocol stack proposed by Wehner et al. \cite{dahlberg19, kozlowski19, kozlowski20} features layers named \textit{physical}, \textit{link}, \textit{network}, \textit{transport}, and \textit{application}, drawing inspiration from classical protocol stacks. The \textit{physical} layer focuses on the hardware components required to generate initial entanglement within a predefined time slot. The \textit{link} layer enhances the robustness of connections between nodes in the quantum internet, leveraging a quantum entanglement generation protocol (QEGP). Notably, this model incorporates key hardware parameters, leading to the development of a \textit{hardware-abstraction} sub-layer that bridges the physical systems to the link layer. The \textit{network} layer utilizes these connections to design network protocols that facilitate endpoint communication across the quantum internet while managing the network's entanglement resources. The \textit{transport} layer is responsible for managing quantum internet traffic, including congestion control \cite{leone21, li23}, re-transmission, and monitoring quantum channel quality. Finally, the \textit{application} layer ensures the functionality of desired applications on quantum devices interconnected by the quantum internet. 

The protocol stack developed by D\"{u}r et al. \cite{pirker18, pirker19} represents a significant departure from previously discussed stacks, as it employs multipartite states instead of the conventional Bell states. This stack includes the layers \textit{physical}, \textit{connectivity}, \textit{link}, and \textit{network}. The \textit{physical} layer in this framework undertakes multiple roles, including the generation and transmission of entangled states, as well as signal conversion between quantum channels. The \textit{connectivity} layer is dedicated to establishing long-distance entanglement through entanglement purification processes. Following this, the \textit{link} layer addresses incoming requests by providing the necessary quantum states. Lastly, the \textit{network} layer ensures the distribution and sharing of entangled states across various networks, thus facilitating broad quantum communication capabilities.

\section{Performance evaluation}\label{sec:performance_evaluation}

In this section, we focus on the evaluation aspects of quantum networks.
In the absence of large-scale field trials, this relies on simulations (Section~\ref{ssec:simulation}) or test bed experiments (Section~\ref{ssec:test_beds}).

\subsection{Simulation}\label{ssec:simulation}
Given the absence of a functional quantum network and the limited experimental capability to manage only a few nodes, as will be discussed in Section~\ref{ssec:test_beds}, the design of quantum networks poses significant challenges, extending to the broader scope of the quantum internet.

First, finding a ``realistic'' network topology is an issue \emph{per se}, which so far has been addressed inspiring by classical networks.

A significant amount of current research utilises the Waxman model \cite{waxman88} to acquire the random topology of the quantum network. Others use lattices (grid and its variations) \cite{vk23, vk24}, rings \& spheres \cite{schoute16}, linear chains \cite{dai20}, and specialised topology, that is, SURFnet \cite{chakraborty20, rabbie22, nguyen22} and US backbone network \cite{li22, nguyen22}.

\begin{figure}
\centering
    \includegraphics[width=\columnwidth]{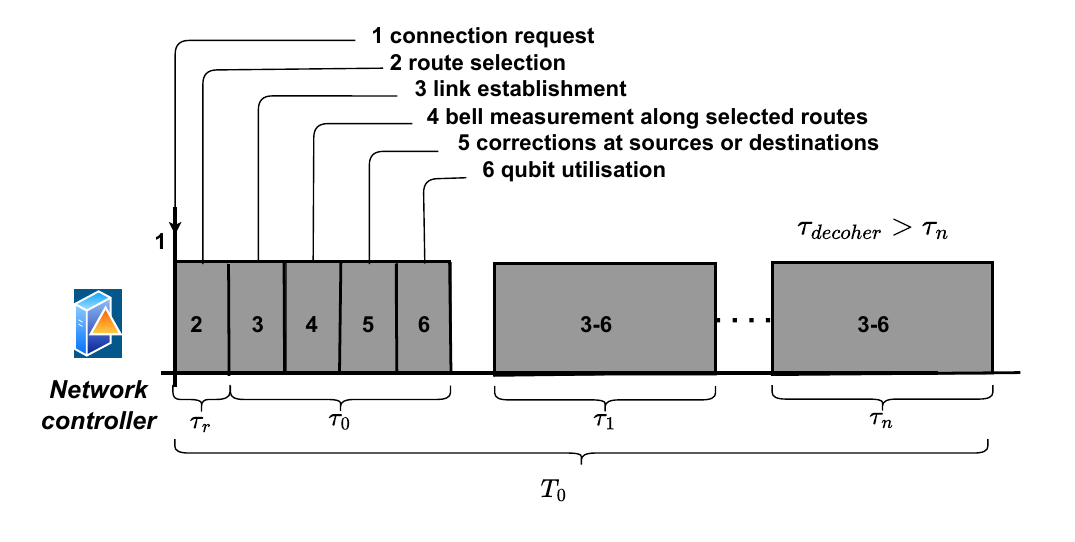}
    \caption{Timing diagram for centralized entanglement distribution. In each period $T_0$, a total of $n+1$ EPR pairs are allocated for a given set of connection requests, with the $i^{th}$ pair distributed in its designated time-slot $\tau_i$. Routes are determined at the start of the period during the $\tau_r$ slot and remain fixed throughout. A new period is initiated whenever a change in the connection requests occurs (modified from \cite{vk24})}
    \label{fig:time_diagram}
\end{figure}

For simplicity, simulators generally employ synchronised time slots $\tau_n \forall \ n \in \mathbf{N}$, which means the node clocks are assumed to be perfectly synchronised, during which the quantum states are presumed to remain coherent, as illustrated in Figure~\ref{fig:time_diagram}. Within this time slot, the distribution of end-to-end entanglement is conducted, that is, the internal and external links establishing the path between the endpoints \cite{pant19, shi20}. The external links are the initial EPR pairs shared between the neighbouring nodes through the quantum channels denoted by block 3 in figure~\ref{fig:time_diagram}. The internal links are within the quantum repeaters to which entanglement swapping is attempted to complete the route between the two endpoints to complete the request denoted collectively by blocks 4 \& 5 in figure~\ref{fig:time_diagram}. This time slot is reiterated until a modification in the connection requests occurs, which may manifest either through the fulfilment of existing requests or the incorporation of new ones \cite{vk24}. All the operations such as quantum gate application performed are usually assumed to be executed in negligible time. 

The commonly referenced synchronised time-slot structure above is a simplified assumption used to organise the already complex protocols for entanglement distribution. However, some recent studies have begun to challenge this assumption by exploring asynchronous structures for routing protocols. For instance, a recent study \cite{yang24} investigates an asynchronous structure that allows for dynamic, distributed updates to network topology using models like DODAG (destination-oriented directed acyclic graph) or spanning tree. Another study proposes a quantum network where nodes autonomously manage the distribution of entanglement to multiple requests asynchronously in a decentralised setting \cite{chen23}. This protocol draws inspiration from OSPF (Open Shortest Path First), which is critical in the classical internet's architecture. These asynchronous approaches, which do not require time synchronisation among nodes, show promise for enhancing the scalability of quantum networks.

As introduced in Section~\ref{ssec:routing_forwarding_scheduling}, routing metrics are required for entanglement distribution. In this regard, a wide range of routing metrics has been used in the literature including throughput \cite{shi20, chakraborty20, zhang21, zhao21}, inverse throughput \cite{meter13}, continuous fidelity curves (entanglement generation fidelity vs rate) \cite{coutinho23}, end-to-end entanglement rate of the path \cite{caleffi17}, hop count \cite{li21}, latency \cite{cicconetti21}, E2E fidelity \cite{zhao22, vk23, vk24}, and fairness \cite{yang22}. The selection of routing paths using the routing metric can also be categorised using different routing algorithms such as Dijkstra-based \cite{meter13, vk23, vk24}, linear programs \cite{iacovelli24}, stochastic programs \cite{kaewpuang23}, integer programming \cite{zeng23}, greedy algorithms \cite{pant19, pandey23}, and AI-based routing algorithms \cite{le22, chaudhary23, islam24}. Additional critical parameters may be constant or variable and encompass link generation success rates and entanglement swapping efficiencies. The model also differentiates between two types of topology information—local \cite{van23} and global \cite{chakraborty19}—which are critical in making routing decisions for end-to-end entanglement. Furthermore, as discussed earlier, the simulation incorporates two modes of heralded entanglement generation, essential for enabling entanglement swapping: on-demand \cite{caleffi17, shi20} and advanced \cite{chakraborty19, li21}, as depicted in figure~\ref{fig:advanced_and_on_demand}.

A majority of current research on the quantum internet is done using simulation tools due to the obvious non-existence of the quantum internet or networks. The notable ones are NetSquid \cite{coopmans21}, SimulaQron \cite{dahlberg18}, QuNetSim \cite{diadamo21}, and SeQUeNCe \cite{wu21}. A list of quantum network simulators can be found in table~\ref{tab:list_sim}. A recent survey on the simulation tools can be found in \cite{bel24}.

\begin{table}[h!]
\centering
\caption{List of quantum network simulators}
\label{tab:list_sim}
\begin{tabular}{p{1.5cm}p{1.5cm}p{1.5cm}p{1.5cm}}
\toprule
Tool & Year & Last updated\textsuperscript{$\dagger$} & Reference \\
\midrule
NetSquid & 2021 & 2021 & \cite{coopmans21} \\
SimulaQron & 2018 & 2021 &  \cite{dahlberg18} \\
QuNetSim & 2021 & 2023 & \cite{diadamo21} \\
SeQUeNCe & 2019 & 2024 & \cite{wu21} \\
SimQN & 2023 & 2023 & \cite{chen23sim} \\
QNET & 2023 & 2023 & \cite{fang23} \\
Squanch & 2018 & 2019 & \cite{bartlett18} \\
\bottomrule
\end{tabular}
\begin{flushleft}
\textsuperscript{$\dagger$} as on August 30, 2024
\end{flushleft}
\end{table}

\subsection{Test beds}\label{ssec:test_beds}

\begin{table}[h!]
\centering
\caption{Summary of entanglement-based quantum network test beds}
\label{tab:summary_entanglement_network}
\begin{tabular}{p{0.6cm}p{2.5cm}p{6.5cm}p{1.6cm}}
\toprule
Year & Number of nodes & Description & Reference \\
\midrule
2021 & $\{4, 5, 8\}$ & Reconfigurable entanglement distributing network with resource-optimised topology & \cite{appas21} \\
2021 & 3 & Entanglement based network with remote solid-state qubit & \cite{pompili21} \\
2021 & 3 & A flex-grid entanglement distribution network & \cite{alshowkan21} \\
2022 & 2 & Entanglement of rubidium atoms with telecom fibre & \cite{van22} \\
2022 & 2 & Post-selected entanglement between atomic ensembles & \cite{luo22} \\
2022 & 3 & Teleportation of a qubit using entanglement swapping with memory storage  & \cite{hermans22} \\
2023 & 2 & Entanglement of trapped-ion qubits using ion-photon entanglement & \cite{krutyanskiy23} \\
2024  & 3 & Memory to memory entanglement with telecom conversion & \cite{liu24} \\
2024 & 2 & Entanglement between nuclear spin memories in a telecom network & \cite{knaut24} \\
2024 & 2 & Automated polarisation-entangled photons distribution over deployed fibre& \cite{craddock24} \\
\bottomrule
\end{tabular}
\end{table}

Due to common quantum mechanics fundamentals, the developments in quantum computing technology have led to trust in the reality of the quantum internet. This has resulted in the emergence of different test beds for quantum networks.

As discussed in Section~\ref{sec:architecture}, the initial simplicity of quantum internet architecture has catalysed a series of QKD network experiments worldwide. These experiments vary by several parameters, including the number of nodes, the type of link used (either fibre optic or free space), the secret key generation rate, the specific QKD protocol employed, and the use of optical switching. Early implementations have typically involved short-range QKD networks and local QKD networks, employing optical components like beam splitters. These foundational networks pave the way for more complex metropolitan and backbone networks. Among these, metropolitan QKD networks have shown the most advancement in experimental implementation. A detailed listing of experimentally implemented QKD-based networks is provided in \cite{cao22}.

The higher stages of quantum internet that require end-to-end entanglement distribution are still in nascent stages. The experimental realisation of these types of networks is currently limited to a few nodes with varying link lengths and physical systems of qubits. A non-exhaustive list of entanglement-based quantum network test beds is provided in the table~\ref{tab:summary_entanglement_network}.

\section{Standardisation}\label{sec:current_initiatives}

Standardisation bodies play a key role in adopting new technology.
For the classical Internet, they have significantly contributed to broadening the ecosystem of industrial players and fostering investments, thanks to smoother interoperability between multi-vendor equipment and a clear definition of the interfaces between the system components.
Due to the technology limitations highlighted so far, the Quantum Internet has not yet attracted sufficient interest to trigger this process, despite the substantial market investments, which happened mostly through the funding of small, yet fast-growing, companies like Aliro Quantum\footnote{\url{https://www.aliroquantum.com/}}, Qunnect\footnote{\url{https://www.qunnect.inc/}}, and QuTech\footnote{\url{https://qutech.nl/}}.
In this respect, efforts so far have been directed at QKD systems, for which technology is more mature, resulting in the Focus Group on Quantum Information Technology for Networks (FG-QIT4N) established at the International Telecommunication Union Telecommunication Standardization Sector (ITU-T), which is now closed and superseded by the Joint Coordination Activity on Quantum Key Distribution Network (JCA-QKDN), and the Industry Study Group (ISG) on QKD at the European Telecommunications Standards Institute (ETSI).

To the best of our knowledge, the only global standardisation initiative also covering more general scenarios is the Quantum Internet Research Group\footnote{\url{https://www.irtf.org/qirg.html}} (QIRG), established within the Internet Research Task Force (IRTF), which is the twin on the Internet Engineering Task Force (IETF) focusing on longer-term research issues.
The group so far has published two information Request For Comment (RFC) documents describing possible use cases and application scenarios for the Quantum Internet~\cite{rfc9583} and laying down some high-level principles for the design of the architecture of the Quantum Internet~\cite{rfc9340}.
Despite the importance of their role in clarifying the motivation and terminology in a tutorial manner and being a source of inspiration to find new research challenges, neither of these documents provides a reference model or interfaces/protocols, which leaves the issue open.

There is one initiative that may fill this gap shortly.
The Quantum Internet Alliance\footnote{\url{https://quantuminternetalliance.org/}} (QIA), funded in 2017, is a European collaboration endeavour with the ambition of fostering an ecosystem ready for the development of the Quantum Internet.
In addition to carrying out research projects aimed at progressing state-of-the-art in specific areas, through joint work of different partners in academy and industry, the alliance is set to provide to the global research community open tools and platforms to speed up research and development and, at the same time, define baselines and reference architectures.
A similar initiative, though at a national level, is QUANT-NET~\cite{monga23}, supported by the U.S. Department of Energy (DOE) under the Advanced Scientific Computing Research (ASCR) program, which is progressing towards the realisation of an open testbed for distributed quantum computing experiments.
Finally, many network infrastructures are being provisioned for a stable operation of QKD backbones, for example at the Oak Ridge National Laboratory in the U.S.~\cite{peters22} (city-level), in ChaQra, the Indian Quantum Network~\cite{gupta24} (national level), and as part of the initiatives of the EuroQCI in Europe\footnote{\url{https://digital-strategy.ec.europa.eu/en/policies/european-quantum-communication-infrastructure-euroqci}} (international level). Other notable efforts include by China (regional/inter-city) Beijing to Shanghai secure network \cite{qiu14}, and ground to satellite QKD \cite{liao17, liao18, chen21km4600}. As the missing pieces of the Quantum Internet become available as commercial devices, it will be possible to install them within such infrastructures by reusing the fibre optic cables and part of the physical-layer communication devices, thus promoting them from QKD-only to higher-generation quantum networks.

\section{Challenges and future directions}\label{sec:challenges_and_future}

As the quantum internet is still in its early development stages, it presents numerous challenges and opportunities for future research. One significant obstacle is the lack of consistent key parameters for physical components, which introduces uncertainty in designing optimized strategies. This variability broadens the potential for exploring effective network design strategies. Additionally, the ongoing exploration of new architectures and frameworks further complicates the situation, as it adds layers of complexity to network development.

In this context, we outline both challenges and future research opportunities. Specifically, we first identify the key technological enablers that are crucial for the design and implementation of quantum networks (Section~\ref{ssec:tech_challenges}). Subsequently, we discuss future research directions that could significantly propel advancements in this field, thereby catalysing the maturation and expansion of the quantum internet (Section~\ref{ssec:future_works}). Finally, we explore the potential for quantum and classical networks to coexist and discuss the interplay between these systems in achieving harmonious integration (Section~\ref{ssec:interplaywithclassical}).

\subsection{Technology enablers}\label{ssec:tech_challenges}

In Section~\ref{sec:qi_fundamentals}, qubits are introduced as mathematical objects with specific properties. However, in practice, qubits can be realised using various physical systems, such as superconducting circuits, trapped ions, quantum dots, photonic qubits, and topological qubits. From the perspective of entanglement distribution, a qubit that travels (e.g., a photon) is referred to as a \textit{flying qubit}, while a qubit stored in a quantum memory is called a \textit{memory qubit}. Both types of qubits are essential for the development of the quantum internet—flying qubits facilitate the distribution of entanglement, while memory qubits, as their name suggests, are used for storing qubits. Even if a consensus is reached on the most effective technology for quantum memory storage, its interaction with flying qubits will be critical for its integration into quantum networks. In this regard, \textbf{quantum transduction}, a process that enables the conversion of quantum signals, is a crucial area of exploration \cite{lauk20}, where no solutions for commercial exploitation exist yet.

Quantum repeaters, as key components in quantum networks, are critical for advancements in the field.

Unfortunately, a fully functional physical \textbf{quantum repeater} has yet to be realised. Currently, a quantum repeater is typically an experimental station that simulates the behaviour of a quantum repeater node. The miniaturisation and advancement of quantum repeaters represent a significant challenge, closely tied to broader developments in quantum technology.

Until these two foundational components are not available as commercial off-the-shelf (COTS) products, it is difficult to assign (tentative) dates to the milestones along the road to the quantum internet, as reported in Section~\ref{ssec:stages}.

\subsection{Research challenges}\label{ssec:future_works}

Realising the quantum internet is a global interdisciplinary process. Therefore, there are several open research challenges associated with the quantum internet from many perspectives and covering heterogeneous areas of expertise.

A key direction for insights into the realisation of the quantum internet is to have more \textbf{experimental test beds}. As discussed in Section~\ref{ssec:test_beds}, current experimental test beds are limited in number and number of nodes. This is due to the size and cost of the nodes. Similar to quantum computing, selecting the right physical system remains a big question for realising the quantum repeater and the quantum devices.

As outlined in Section~\ref{ssec:protocol_stack}, while various quantum internet \textbf{protocol stacks} are discussed in the literature, a consensus on the reference model has not yet been established. The lack of a universally superior physical system highlights the need for a well-defined interface between the foundational hardware and the software layers above. Observing the experimental realisation of these proposed stacks is crucial. Progress in this area includes the experimental implementation of link and physical layer protocols within a network consisting of just two nodes based on diamond NV centres \cite{pompili22stackexp}, where entangled states were delivered with the fidelity specified by the user.

Additionally, the unique properties of the quantum internet challenge the suitability of the universal layer model that has been effective in the classical internet. This ongoing debate emphasises the necessity for a customised approach tailored to the specific needs and capabilities of quantum communication technologies.

As detailed in Section~\ref{ssec:routing_forwarding_scheduling}, the distribution of end-to-end entanglement, essential for realising the quantum internet, presents a significant unresolved challenge. Various strategies addressing routing, forwarding, and scheduling have been investigated, yet a comprehensive understanding remains elusive. 
There is currently no definitive model that fully addresses all the complexities of the optimisation problem involved in \textbf{quantum entanglement distribution}, a problem which has no direct classical counterpart. A significant challenge within this model is the inherent non-determinism; there is a probability of failure in both the establishment of EPR pairs used for the swapping procedure and in the entanglement purification process. To address these challenges, a \textit{cost-vector formalism} model has been proposed \cite{leone21} recently. The essence of this model is to characterise each EPR pair with two probability weights: the \textit{transmission probability}, which reflects the likelihood of successfully establishing an EPR pair, and the \textit{coherence probability}, which indicates the likelihood that an EPR pair will remain coherent and thus suitable for distribution to endpoints. Further exploration in this area is necessary to more accurately model and optimise the complex processes involved in routing and distributing entanglement within quantum networks. Furthermore, the \textbf{time-slot structure} proposed in existing literature, which is crucial for maintaining qubit coherence, hinges on the choice of the physical system used for quantum memory. This choice is itself a significant area of research, reflecting the intricate dependencies and innovations required in the development of quantum internet technologies. 

As the focus of research transitions from individual quantum networks to the realisation of a global quantum internet, scalability becomes a critical concern. Current efforts largely concentrate on quantum network designs that incorporate networks of directly connected quantum repeaters to facilitate entanglement distribution. However, the \textbf{scalability} of these networks remains largely unexplored, particularly in scenarios that mirror the complex, multi-service provider environment of the classical internet \cite{liu22multi}. Addressing this, a recent study has proposed an inter-domain routing protocol suitable for a decentralised setting \cite{liu24bgp}. This protocol introduces a novel metric known as \textit{information gain}, which quantifies the amount of entanglement information to aid in the selection of high-fidelity paths. This development adds another layer of complexity, considering the existing challenges in quantum network scalability and functionality.  

Given the fragility of quantum states, \textbf{entanglement purification and quantum error correction} techniques are vital for addressing the challenges inherent in the quantum internet \cite{balapurification25, huroutingerrorcorrection24}. Quantum error correction, in particular, is a broad field of research that intersects significantly with the development of robust quantum computing systems. However, these techniques require additional resource consumption, which presents a trade-off. Ongoing research is focused on optimising the integration of these techniques within the network and application layer protocols of the quantum internet. This research aims to enhance the reliability and efficiency of quantum communications while managing the costs associated with these advanced error mitigation strategies.

Analogous to the crucial role of network addressing in classical networks, which facilitates communication and data exchange, there is a need to explore \textbf{addressing schemes} for quantum networks. This exploration is essential to ensure that quantum networks can effectively manage and route quantum information, similar to how IP addresses function in traditional networks. Developing robust addressing schemes will be key to the successful implementation and scalability of quantum internet technologies.

A significantly understudied area within quantum networking is the development of application interfaces that provide functionalities to the end node. It is reasonable to anticipate the creation of \textbf{Application Programming Interfaces} (APIs) for networked applications specifically tailored for quantum networks. These APIs would be designed to meet the unique requirements of quantum data transmission and processing, enabling efficient, secure, and scalable quantum communication systems.

The development of \textbf{simulation tools} for quantum networking is a critical and ongoing effort, as discussed in Section~\ref{ssec:simulation}, supporting the expansive research in the field of the quantum internet. Recently, various simulation tools have emerged, each focusing on different aspects of modelling. For instance, Netsquid is designed to integrate with high-performance computing environments, SeQUeNCe emphasises user-friendliness, and SimQN offers support for user-defined noise models, among others. The effort to develop and refine these tools represents a substantial area of research and development, highlighting the diverse needs and challenges within quantum network simulation.

As mentioned briefly in Section~\ref{sec:qi_fundamentals}, \textbf{multi-partite entanglement}, as covered by D\"{u}r et al.'s protocol stack, can be employed in quantum networks to achieve the same objectives traditionally aimed with bi-partite entanglement. This area, particularly in comparison to bi-partite entanglement, is significantly understudied and warrants further research \cite{mccutcheonmulti16, avismulti23, ainleymulti24}. A notable feature of entanglement generation using GHZ protocols is the presence of a supercritical region where the entanglement generation rate remains constant regardless of distance \cite{patil22distanceindependent, patil21multiplexedghz}. This contrasts sharply with the exponential decay in entanglement generation rate associated with bell-state protocols, which occurs even with multi-path routing given that the bell measurement success probability $q<1$. 

\subsection{Interplay with classical networks}\label{ssec:interplaywithclassical}
Given the inherent differences in how quantum and classical networks operate, achieving their coexistence on a shared physical infrastructure—as outlined in Section~\ref{sec:intro}—poses a significant challenge. One immediate issue is the simultaneous transmission of quantum and classical signals over the same optical fibers. While classical systems rely on high-intensity signals supported by amplifiers and multiplexing schemes, quantum signals—often at the single-photon level—are extremely weak and highly susceptible to noise such as Raman scattering from adjacent classical channels \cite{Zavitsanosquantumclassicalsignal19}. To protect quantum channels from such interference without compromising classical performance, advanced optical filtering and the development of low-noise components are essential. Moreover, quantum protocols demand synchronization with precisions down to the nanosecond or even picosecond level—requirements that far exceed those of typical classical networks \cite{lukenshybrid25}. Experimental setups have adopted protocols like the White Rabbit system to achieve the necessary synchronization \cite{whiterabbit09}, yet integrating these high-precision timing solutions into existing classical infrastructures remains a formidable task. Additionally, quantum networks inherently utilize two parallel channels: a quantum channel for transmitting qubits and a classical channel for control, coordination, and supplementary data. Integrating these channels requires managing a dual protocol stack, where the classical control plane must orchestrate complex quantum operations such as entanglement generation and swapping \cite{lukenshybrid25, rijsman23}. This duality further increases network management complexity, necessitating the development of novel control algorithms and standardized interfaces for efficient resource allocation, scheduling, and routing across both domains.

\section{Conclusion}\label{sec:conclusion}
In this paper, we present a comprehensive introduction to the fundamental concepts of the quantum internet, distinguishing it from the more established field of quantum computing. We explore the identified applications and use cases, categorising them based on the advantages they offer over existing technologies. A concise discussion follows on the stages of the quantum internet's evolution, its various types, the challenges of entanglement distribution, and the protocol stack. We also review the current landscape of performance evaluation in quantum networks, with a focus on simulation methods, test beds, and the role of standardisation bodies in this field. Finally, we address the present challenges and outline future research directions that will be essential for driving further advancements in the quantum internet.

There has been a clear surge of interest in quantum networks, particularly in recent years. This is largely due to the fact that the initial stages of quantum network implementation face fewer technological hurdles compared to quantum computing, although the challenges for the later stages overlap with those of quantum computing. However, the momentum in this field has primarily been driven by the establishment of standardisation bodies and contributions from small-scale industries, with limited involvement from major industry players. We believe that the key to realising the full vision of the quantum internet lies with the technological enablers, like the quantum repeater and matter/flying qubit transducers. Given the vast solution space created by the current lack of critical physical components, there is a pressing need for sustained research and development efforts to address these gaps and advance the field further.

\backmatter

\section*{Declarations}

\begin{itemize}
\item \textbf{Funding:}
The work of C.~Cicconetti was funded by the European Union -- Next Generation EU under the Italian National Recovery and Resilience Plan (NRRP), Mission 4, Component 2, Investment 1.3, CUP B53C22003950001, partnership on ``SEcurity and RIghts in the CyberSpace'' (PE00000014 ``SERICS'', Spoke 9). 
        The work of M.~Conti and A.~Passarella was funded by the European Union -- Next Generation EU under the Italian National Recovery and Resilience Plan (NRRP), Mission 4, Component 2, Investment 1.3, CUP B53C22003970001, partnership on ``Telecommunications of the Future'' (PE00000001 ``RESTART'', Spoke 1).
        The work was partly funded by the European Union under the project EuroQCI QUID (GA no. 101091408). \\
\item \textbf{Conflict of interest/Competing interests:} The authors declare that they have no competing interests. \\
\item \textbf{Ethics approval and consent to participate:} Not applicable \\
\item \textbf{Consent for publication:} Not applicable \\
\item \textbf{Data availability:} Not applicable \\
\item \textbf{Materials availability:} Not applicable \\
\item \textbf{Code availability:} Not applicable  \\
\item \textbf{Acknowledgements:} Not applicable \\
\item \textbf{Author contribution:} All authors contributed to the initial concept and design of the paper, including drafting the table of contents. VK wrote the first draft of the manuscript and was also responsible for editing and reviewing the final version. CC, MC, and AP each reviewed and edited the manuscript individually. All authors read and approved the final manuscript.
\end{itemize}

\section*{Orcid}
\textit{\textbf{Vinay Kumar}}\orcidlink{0000-0002-4635-3237} 
\href{https://orcid.org/0000-0002-4635-3237}{https://orcid.org/0000-0002-4635-3237} \\
\textit{\textbf{Claudio Cicconetti}}\orcidlink{https://orcid.org/0000-0003-4503-4223} \href{https://orcid.org/0000-0003-4503-4223}{https://orcid.org/0000-0003-4503-4223} \\
\textit{\textbf{Marco Conti}}\orcidlink{https://orcid.org/0000-0003-4097-4064} \href{https://orcid.org/0000-0003-4097-4064}{https://orcid.org/0000-0003-4097-4064} \\
\textit{\textbf{Andrea Passarella}}\orcidlink{https://orcid.org/0000-0002-1694-612X} \href{https://orcid.org/0000-0002-1694-612X}{https://orcid.org/0000-0002-1694-612X} \\

\end{document}